\documentclass[preprint]{aastex63}

\begin{document}

\title{Atmospheric Overturning Circulation on Dry, Tidally-locked Rocky Planets is Mainly Driven by Radiative Cooling}

\author[0000-0002-9406-1781]{Shuang Wang}
\affiliation{Laboratory for Climate and Ocean-Atmosphere Studies, Dept. of Atmospheric and Oceanic Sciences, School of Physics, Peking University, Beijing 100871, China}

\author[0000-0001-6031-2485]{Jun Yang}
\affiliation{Laboratory for Climate and Ocean-Atmosphere Studies, Dept. of Atmospheric and Oceanic Sciences, School of Physics, Peking University, Beijing 100871, China}


\begin{abstract}


In this study, we examine the driving mechanism for the atmospheric overturning circulation on dry, tidally-locked rocky planets without the condensation of water vapor or other species. We find that the main driving process is the radiative cooling of CO$_2$ (or other non-condensable greenhouse gases) rather than CO$_2$ greenhouse warming or stellar radiation. Stellar radiation is the ultimate mechanism but not the direct mechanism. Due to the combination of the uneven distribution in the stellar radiation and effective horizontal energy transports in the free troposphere, there is strong temperature inversion in the area away from the substellar region. This inversion makes CO$_2$ to have a radiative cooling effect rather than a radiative warming effect for the atmosphere, same as that in the stratosphere of Earth’s atmosphere. This cooling effect produces negative buoyancy and drives large-scale downwelling, supporting the formation of a global-scale overturning circulation. If CO$_2$ is excluded from the atmosphere, the overturning circulation becomes very weak, regardless the level of stellar radiation. This mechanism is completely different from that for the atmospheric overturning circulation on Earth or on moist, tidally-locked rocky planets, where latent heat release and/or baroclinic instability are the dominated mechanisms. Our study improves the understanding of the atmospheric circulation on tidally-locked exoplanets and also on other dry planets, such as Venus and Mars in the solar system.

\end{abstract}


\section{Introduction} \label{sec:intro}

What drive the atmospheric circulations of different rocky planets that have various atmospheric compositions? For planets without strong geothermal or tidal heating, atmospheric circulations are ultimately driven by stellar energy and its spatial distribution, but the intermediate processes connecting the stellar energy to the atmospheric circulations could be completely different for different planets. In this study, we plan to dig out the driving mechanism(s) for the atmospheric overturning circulation on dry, tidally-locked rocky planets around low-mass stars. The atmospheric compositions are assumed to be dry without water vapor condensation or the condensation of any other species.

\begin{figure}[h]
    \centering
    \includegraphics[width=\textwidth]{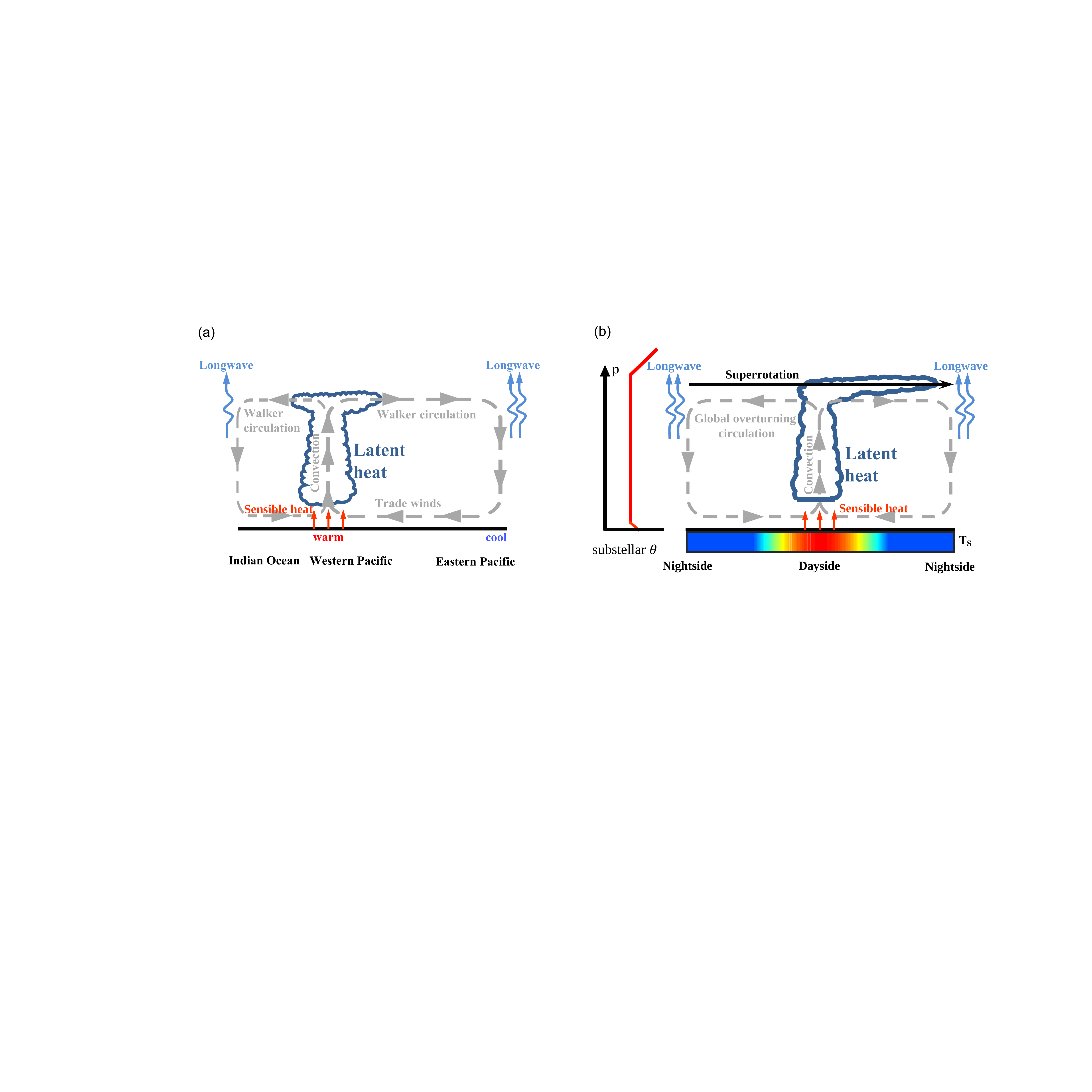}
    \caption{Schematic diagrams of \textcolor{black}{(a) long-term mean tropical Walker circulation on Earth} and (b) atmospheric circulation on moist, tidally-locked rocky planets. Grey arrows represent overturning circulations, and black arrow in panel (b) represents equatorial superrotation. The red and blue arrows represent sensible heat and longwave radiation, respectively. Clouds form over the convective regions and are blown by upper winds. During cloud formation, latent heat releases, and it promotes the strong overturning circulation. In panel (b), the potential temperature profile in substellar region (red line on the left) shows isentropic structure in the troposphere, indicating the range of the deep convection.}
    \label{fig:clima_diag}
\end{figure}

On Earth, latent heat release in the deep tropics and longwave radiative cooling in the subtropics \textcolor{black}{in addition to the effect of planetary rotation} support the upwelling and downwelling branches of the tropical \textcolor{black}{mean meridional} circulation--the Hadley cells \citep{Holton_2013,Vallis_2017}. Numerical simulations showed that the Hadley cells become very weak if latent heat release was turned off on Earth \citep{Frierson_2006,caballero_2008,Horinouchi_2012,Li_juan_2020,Suhas_2021}. In the simulations, three different methods have been employed to make the atmosphere be dry, turning off the surface evaporation, eliminating the latent heat release of fusion and vaporization, and changing the Clausius--Clapeyron relation. \textcolor{black}{Meanwhile, latent heat release mainly over the western Pacific and longwave radiative cooling over the eastern Pacific drive the tropical Walker circulation (Figure~\ref{fig:clima_diag}(a)), we well disscuss this more in section~\ref{sec:results_3}.} In the middle latitudes, the atmospheric \textcolor{black}{mean meridional} circulation is mainly driven by baroclinic instabilities related to the steep meridional (south--north) temperature gradients. Baroclinic eddies cause eddy heat flux divergences in the subtropics and eddy heat flux convergences in middle latitudes, leading to zonal-mean descent in the subtropics and ascent in higher latitudes and to form an indirect Ferrel circulation on each hemisphere \citep{Vallis_2017}. The Hadley cells are connected to the Ferrel cells by baroclinic eddies that propagate from the mid-latitudes towards the tropics and meanwhile transport momentum from the tropics to extratropics.

Atmospheric circulation on moist, slowly rotating, tidally-locked rocky planets has been simulated using global general circulation models (GCMs) in many previous studies \citep[e.g.,][]{Merlis_2010,Edson_2011,Leconte_2013,Yang_2013,Kopparapu_2016,Kopparapu_2017,Noda_2017,Haqq_Misra_2018,Ding_2020_water-rich}. The simulation results agree that the atmospheric circulation is characterized by an equatorial superrotation, wavenumber-1 stationary planetary wave, and a nearly isotropic overturning circulation consisting of robust upwelling over the substellar region and moderate subsidence in the rest of the planet, as shown in \textcolor{black}{Figure~\ref{fig:clima_diag}(b)}. The overturning circulation is an analog for the \textcolor{black}{Walker circulation} on Earth but extends to the whole planet. The driving mechanisms are latent heat release over the substellar region and longwave radiative cooling in the rest of the planet, same as those in the tropics of Earth. For moist, rapidly rotating, tidally-locked rocky planets, the atmospheric circulation is more complex because of the onset of baroclinic instabilities in high latitudes, and meanwhile the equatorial superrotation is stronger due to the fast rotation rate and more effective equatorward momentum transports \citep{Merlis_2010,Noda_2017,Haqq_Misra_2018,Pierrehumbert_Hammond_2019,Sergeev_2021}.

For dry, tidally-locked rocky planets, GCM simulations showed that there is a strong atmospheric overturning circulation, and its strength is comparable to that in a moist atmosphere \citep{Joshi_1997,Edson_2011,Leconte_2013,Wordsworth_2015,Koll_2016,Turbet_2021}. These articles assumed that the underlying mechanism is the uneven distribution of stellar radiation and the large surface temperature difference between dayside and nightside (this is the ultimate mechanism, but the intermediate processes are unclear). However, a recent study of \cite{Ding_2019} showed that the atmospheric overturning circulation becomes stronger when CO$_2$ is added in their experiment, within which the surface temperature contrast between dayside and nightside becomes smaller (see their Figures 7 and 8). This result implies that surface temperature contrast may not be the direct driving mechanism, and other mechanism(s) should play an important role.

The goal of this study is to improve the understanding of the atmospheric overturning circulation on dry, tidally-locked rocky planets, based on 3D GCM simulations. We confirm that the CO$_2$ radiative cooling in a dry atmosphere replaces the role of latent heat release in a moist atmosphere, and it connects the uneven distribution of stellar energy and suface temperature to the atmospheric overturning circulation. We describe the models and experimental designs in Section~\ref{sec:methods}. Section~\ref{sec:results} shows the greenhouse warming and radiative cooling effects of CO$_2$ (Section~\ref{sec:results_1}), the atmospheric circulation (Section~\ref{sec:results_2}), and the underlying mechanism (Sections~\ref{sec:results_3} and~\ref{sec:results_4}). Section~\ref{sec:summary} is the summary.

\section{Model Descriptions and Experimental Designs}\label{sec:methods}
For most of the results presented below, we use the Community Atmosphere Model version 3.1 \citep[CAM3;][]{Collins_2004}. CAM3 solves the primitive equations for atmospheric motion on a rotating sphere and the equations for radiative transfer including the effects of water vapor, greenhouse gases, and clouds. The model supports three different dynamical cores of Eulerian, semi-Lagrangian, and finite volume, and here we use the Eulerian core. Sub-scale physical processes are parameterized. For example, deep convection is parameterized based on the Zhang-McFarlane convection scheme \citep{Zhang_1995}, cloud fraction scheme is introduced based on \cite{Slingo_1987}, and boundary turbulent mixing and diffusion are parameterized based on \cite{Byron_2003}. We run some additional experiments using CAM version 4 \citep[CAM4;][]{Neale_2010} and Exoplanet CAM\footnote{https://github.com/storyofthewolf/ExoCAM} (ExoCAM). CAM4 has a new deep-convection scheme relative to CAM3 \citep{Raymond_1986,Richter_2008}. ExoCAM is based on CAM4 but with an improved radiative transfer module modified by Dr. Eric Wolf \citep{Wolf_2014,Wolf_2015,Wolf_2017,Wolf_et_al_2017,Wolf_2022}. The new radiative transfer module uses the correlated-\emph{k} method \citep{Mlawer_1997}, and the correlated-\emph{k} coefficients are calculated with access to the HITRAN 2004 spectroscopic database \citep{Rothman_2005}. The radiative transfer module is more accurate than CAM3 and CAM4 when CO$_2$ concentration or water vapor concentration is much higher than modern Earth's levels (although not employed in this study). Besides, we use the dynamical core of the finite volume method for CAM4 and ExoCAM.

By default, the three models are run for moist atmospheres, and we have modified them to be applicable to simulate completely dry atmospheres. We turn off the moisture fluxes between the atmosphere and the surface, and remove all clouds and water vapor in the initial fields. These models are coupled to a surface layer holding a thermal inertia being equal to 1-meter-depth water. The surface albedo is set to be 0.2 everywhere.

The M-type stellar spectrum data for AD Leo \citep{Segura_2005} is used. Planetary radius is 1.5 $R_{\oplus}$ ($R_{\oplus}$ is Earth's radius), and the gravity is 13.5 $\rm{m\,s^{-2}}$. By default, planetary rotation period (= orbital period) is set to be 60 Earth days, with assuming that the planet is in 1:1 tidally-locked or called synchronously rotating. Additional experiments with rotation period (= orbital period) of 5 Earth days are also examined, to know whether the conclusions are applicable to rapidly rotating planets. The background atmosphere is N$_2$ of 1.0 bar. CO$_2$ is the only greenhouse gas in our experiments. CAM3 and CAM4 do not calculate N$_2$-N$_2$ collision-induced absorption, so that a pure N$_2$ atmosphere is transparent to both stellar and terrestrial radiation. By default N$_2$-N$_2$ collision-induced absorption is included in ExoCAM, but we turn it off to ensure the pure N$_2$ atmosphere is transparent to radiation. So that, we could examine the separated effect of CO$_2$ radiation. However, the pressure broadening effect of N$_2$ on the thermal absorption of CO$_2$ is included.

Two key experiments are designed, one is with no CO$_2$ (actually the CO$_2$ concentration is set to 10$^{-30}$ ppmv because the model does not allow a zero value) and the other one is with 100-ppmv CO$_2$. Below, we call the first scenario as ``pure N$_2$'' and the second scenario as ``N$_2$+CO$_2$''. Besides of these two cases, a series of other CO$_2$ concentrations is also tested, 10$^{-6}$, 10$^{-5}$, 10$^{-4}$, 10$^{-3}$, 10$^{-2}$, 10$^{-1}$, and 1 ppmv. By default, the stellar flux at the substellar point is 1200 W\,m$^{-2}$. In order to test the effect of insolation strength, a series of stellar flux is tested, 1400, 1600, 1800, 2000, and 3000 W\,m$^{-2}$. For each stellar flux, the two scenarios, ``pure N$_2$'' and ``N$_2$+CO$_2$'', are run.

We run all CAM3 simulations with a spatial resolution of $3.75^{\circ}\times3.75^{\circ}$ in latitude and longitude for horizontal grids and 26 inhomogeneous levels for vertical grids, and run CAM4 and ExoCAM with $4^{\circ}\times5^{\circ}$ and 40 vertical inhomogeneous levels. We reduce the model time step from its standard value of 1200 s to 300 s for numerical stability, because the surface temperature at the substellar point could be very high, such as 370 K, on a dry planet without surface evaporation and clouds. The phase velocities of gravity waves are proportional to the square root of air temperature, so that a dry atmosphere is more likely to violate the Courant–Friedrichs–Lewy criterion \citep{Koll_2015}.

In the following text, to better illustrate the thermal structure and atmospheric overturning circulation from dayside to nightside, we use the tidally-locked (TL) coordinates\footnote{https://github.com/ddbkoll/tidally-locked-coordinates} \citep[ref.][]{Koll_2015}. For the latitudes in the TL coordinates, 90$^{\circ}$ is the substellar point (SP), $-$90$^{\circ}$ is the antistellar point (AP), and 0$^{\circ}$ is the terminator. However, TL coordinates cannot well show equatorial superrotation, so we still use standard spherical coordinates in discussing the equatorial superrotation as well as zonal-mean zonal winds.

\section{Results}\label{sec:results}

\subsection{Thermal structure}\label{sec:results_1}
\begin{figure*}
    \centering
    \includegraphics[width=0.7\textwidth]{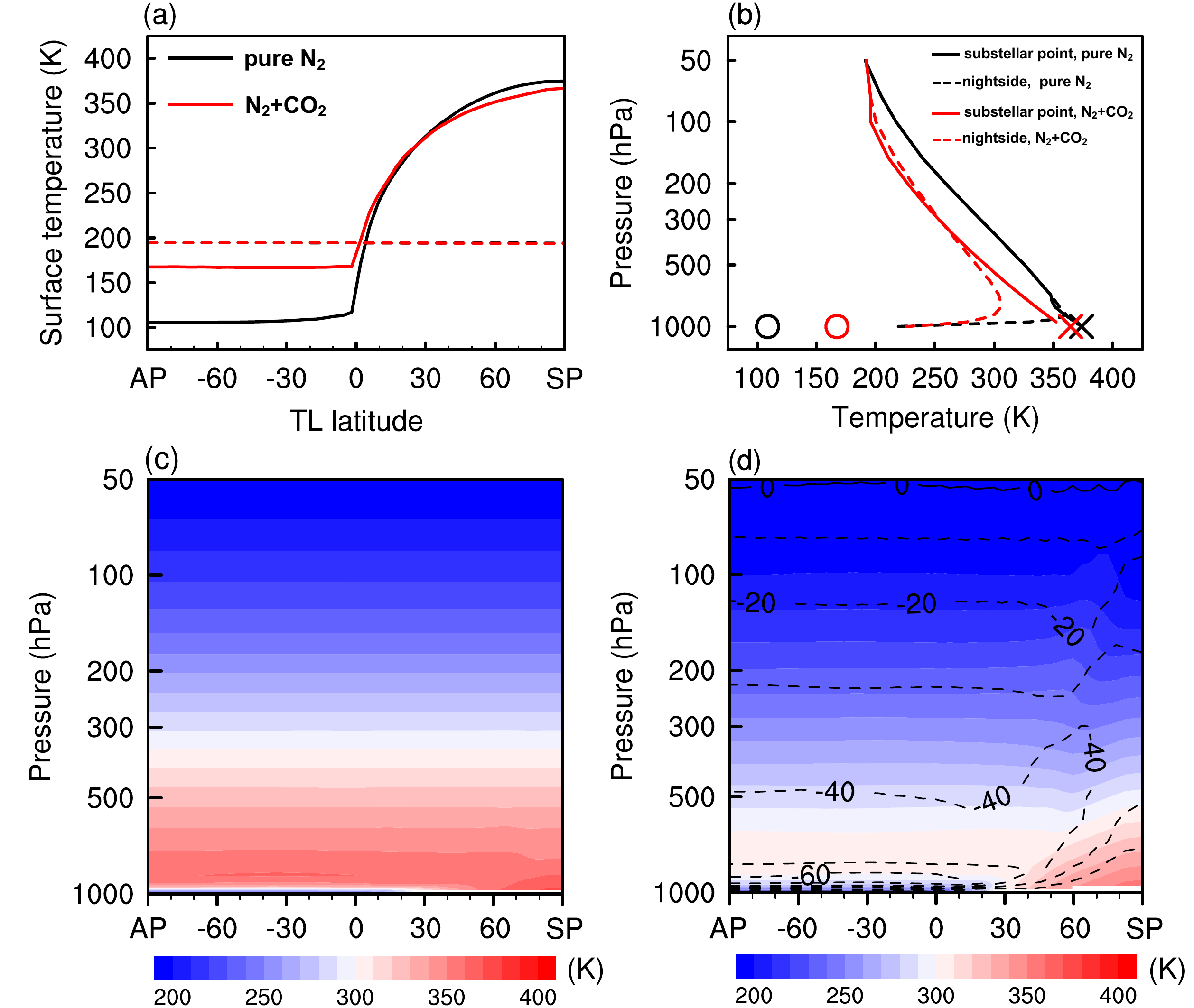}
    \caption{Thermal structures in the CAM3 experiments. (a) Surface temperatures in the pure N$_2$ (black solid) and N$_2$+CO$_2$ (red solid) experiments, and the red dashed line represents CO$_2$ condensation temperature at 1.0 bar. (b) Substellar-point (solid) and nightside-averaged (dashed) temperature profiles. The substellar-point and nightside-averaged surface temperatures are marked by crosses and circles, respectively. (c) Temperature structure in the pure N$_2$ experiment. (d) Same as (c) but in the N$_2$+CO$_2$ experiment, with dashed lines representing the differences from (c) (in units of K). Panels (a), (c), and (d) are shown in tidally-locked coordinates and are averaged over the tidally-locked longtitude. AP is the antistellar point, and SP is the substellar point.}
    \label{fig:MAIN_surf-vert_temp}
\end{figure*}

The surface temperature distribution in the CAM3 experiment of a pure N$_2$ atmosphere is shown in Figure~\ref{fig:MAIN_surf-vert_temp}(a). Without clouds and water evaporation, the dayside surface is very hot. The hottest region is at the substellar point, about 375 K. The dayside surface temperature along the tidally-locked latitudes decreases nearly as a cosine function, analogous to the incoming stellar flux. Without exposure to the star, the nightside surface temperature is nearly uniform and very low, only 105 K. Figure~\ref{fig:MAIN_surf-vert_temp}(b) shows its nightside-averaged and substellar-point vertical temperature profiles. They are overlapped in almost the whole troposphere except close to the surface, which indicates uniform horizontal temperature distributions and hence a weak temperature gradient (WTG) regime in the free atmosphere (Figure~\ref{fig:MAIN_surf-vert_temp}(c)), due to the slow rotation. The nightside-averaged profile shows a strong temperature inversion near the surface, and the near-surface atmosphere on the nightside is about 100 K warmer than the surface. Figure~\ref{fig:MAIN_surf-vert_temp}(c) indicates that the horizontal range of the temperature inversion extends to the dayside. This temperature inversion is due to the combination of the uneven distribution in the stellar radiation and the effective horizontal atmospheric energy transport especially in the free troposphere. A similar inversion layer can be found in the stratosphere or in the polar boundary layer on Earth \citep{Joshi_2020}.

Figures~\ref{fig:MAIN_surf-vert_temp}(a) and (b) show that adding 100-ppmv CO$_2$ to the atmosphere raises the night-side surface temperature by about 60 K, due to the greenhouse effect of CO$_2$. However, surface temperature in the substellar region shows a decrease, about 8 K. Previous studies of \cite{Kite_2011}, \cite{Wordsworth_2015}, and \cite{Koll_2016} showed similar phenomena. For example, \cite{Kite_2011} found that the surface temperature generally increases when the surface air pressure is increased, except in the substellar region the surface temperature decreases (see their Figure 2). \cite{Wordsworth_2015} found that when the pressure of CO$_2$ is increased from 0.01 to 1 bar, the overall surface temperature increases but the substellar surface temperature decreases (see their Figure 2). This phenomenon is due to increased energy transport from dayside to nightside. In our experiment of 100-ppmv CO$_2$, the energy transport dramatically enhances from 8 to 71 W\,m$^{-2}$, leading to the cooling of the substellar surface. \textcolor{black}{The efficient energy transport is due to the stronger day-night overturning circulation, which will be shown in the subsequent section.}

Figures~\ref{fig:MAIN_surf-vert_temp}(b) and (d) also show that adding CO$_2$ obviously cools the whole free atmosphere. For example, the air temperature at the substellar point decreases by 16 K at 800 hPa, and the decreasing of the nightside air temperature is 20--60 K. The mechanisms for the substellar atmospheric cooling and for the nightside free tropospheric cooling are different. The nightside tropospheric cooling is similar to the cooling in Earth's stratosphere as increasing CO$_2$ concentration \citep[][etc.]{Manabe_1967,Langematz_2003,Shine_2003,Fomichev_2007}, because they have a similar thermal structure---temperature inversion. When CO$_2$ is added to the inversion layer, atmospheric emissivity increases, which acts to cool the air there. The decrease of air temperature in the substellar region is due to more horizontal heat transport from dayside to nightside.

\begin{figure*}
    \centering
    \includegraphics[width=0.7\textwidth]{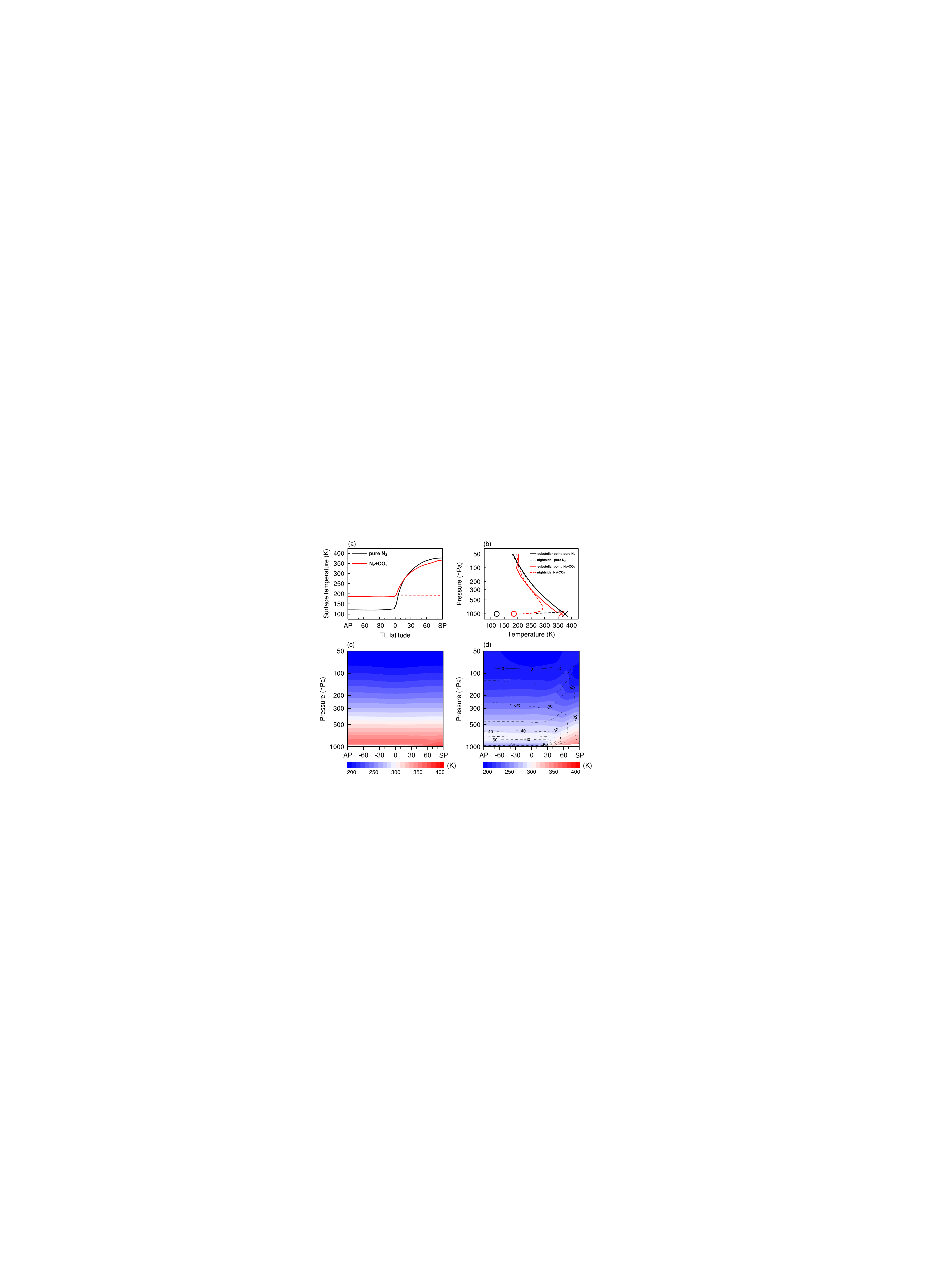}
    \caption{Same as Figure~\ref{fig:MAIN_surf-vert_temp} but for the results of ExoCAM experiments.}
    \label{fig:MAIN_ExoCAM}
\end{figure*}

Figure~\ref{fig:MAIN_ExoCAM} shows the thermal structure simulated by ExoCAM, giving an analog for results of CAM3. The surface temperature for the pure N$_2$ atmosphere is 377 K at the substellar point but 120 K on the nightside. Vertical thermal structures show weak temperature gradients in the free troposphere although large day-night contrasts near the surface. There is also obvious temperature inversion away from the substellar point. When adding CO$_2$, ExoCAM has similar tendencies to CAM3, including about 65 K warming at the nightside surface and about 10 K cooling at the substellar surface. Figures~\ref{fig:MAIN_ExoCAM}(b) and (d) show that adding CO$_2$ cools the whole atmosphere above the boundary layer, same as CAM3.

Why does the surface warm but the overlying atmosphere cool when the CO$_2$ concentration is increased? We employ a simple one-layer atmosphere model with two columns representing the dayside and nightside to understand this phenomenon, similar to \cite{Ding_2019}. In the column model, the dayside and nightside surface are blackbodies with temperatures of $T_{s,d}$ and $T_{s,n}$, the respective overlying atmosphere has temperatures of $T_{a,d}$ and $T_{a,n}$, and the atmospheric emissivity is $\epsilon$. The net stellar flux on the dayside is $S^*$ while it is zero on the nightside, and the atmosphere is transparent to stellar radiation. Ignoring surface sensible heat, energy budgets for the surface and the atmosphere are
\begin{equation}\label{eq:energy_budget_dayside_surf}
     \sigma T_{s,d}^4=\epsilon\sigma T_{a,d}^4+S^*,
\end{equation}
\begin{equation}\label{eq:energy_budget_dayside_atmos}
    \epsilon\sigma T_{s,d}^4=2\epsilon\sigma T_{a,d}^4+C,
\end{equation}
on the dayside, and
\begin{equation}\label{eq:energy_budget_nightside_surf}
     \sigma T_{s,n}^4=\epsilon\sigma T_{a,n}^4,
\end{equation}
\begin{equation}\label{eq:energy_budget_nightside_atmos}
    \epsilon\sigma T_{s,n}^4+C=2\epsilon\sigma T_{a,n}^4,
\end{equation}
on the nightside, where $C$ ($C>0$) is atmospheric energy transport from dayside to nightside. Combining Equations~(\ref{eq:energy_budget_dayside_surf})--(\ref{eq:energy_budget_nightside_atmos}) yields,
\begin{equation}\label{eq:T_sd}
    T_{s,d}=\left[\frac{2S^*-C}{\sigma(2-\epsilon)}\right]^{1/4},
\end{equation}
\begin{equation}\label{eq:T_ad}
    T_{a,d}=\left[\frac{S^*}{\sigma(2-\epsilon)}-\frac{C}{\sigma\epsilon(2-\epsilon)}\right]^{1/4},
\end{equation}
\begin{equation}\label{eq:T_sn}
    T_{s,n}=\left[\frac{C}{\sigma(2-\epsilon)}\right]^{1/4},
\end{equation}
\begin{equation}\label{eq:T_an}
    T_{a,n}=\left[\frac{C}{\sigma\epsilon(2-\epsilon)}\right]^{1/4}.
\end{equation}
Equations~(\ref{eq:T_sn}) and (\ref{eq:T_an}) show that for a gray atmosphere ($0<\epsilon<1$), $T_{s,n}$ is smaller than $T_{a,n}$, so temperature inversion is inevitable on the nightside. For a given value of $C$, Equations~(\ref{eq:T_sn}) and~(\ref{eq:T_an}) show that a larger $\epsilon$ (here it is induced by adding CO$_2$) would warm the night-side surface ($T_{s,n}$) but cool the night-side overlying atmosphere ($T_{a,n}$). This is similar to the results shown in Figure 10 of \cite{Koll_2016} and in Figure 7 of \cite{Ding_2019}. Equations~(\ref{eq:T_sd}) and (\ref{eq:T_ad}) suggest that the increased $\epsilon$ tends to warm both the dayside surface ($T_{s,d}$) and the overlying atmosphere ($T_{a,d}$), while the GCMs' results are opposite. This is mainly because the increase of atmospheric energy transport ($C$) from dayside to nightside in the GCMs, as discussed above.

\subsection{Atmospheric overturning circulation}\label{sec:results_2}
\begin{figure*}
    \centering
    \includegraphics[width=0.6\textwidth]{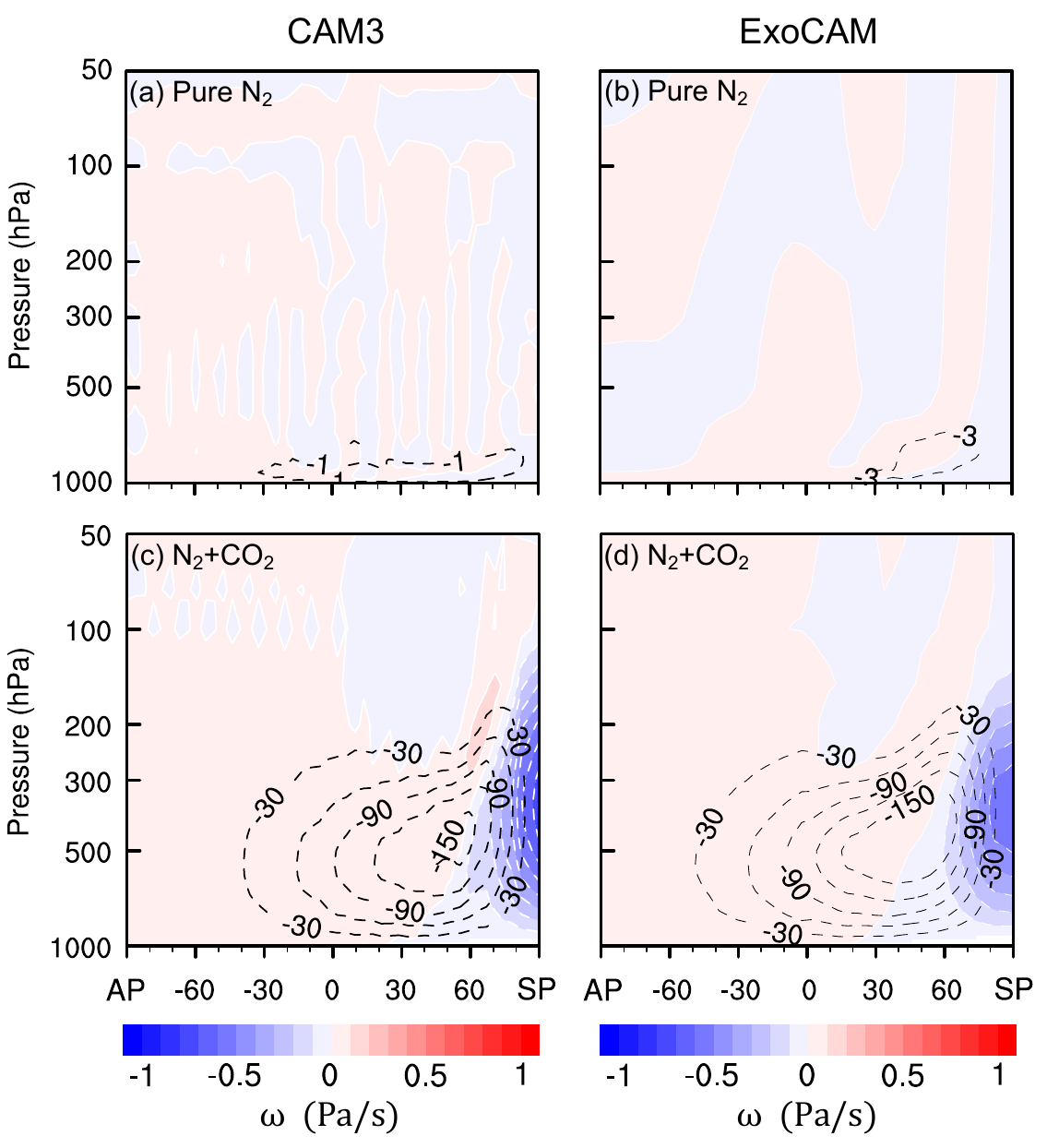}
    \caption{Vertical velocities in pressure coordinates (color shading, negative: upwelling, and positive: downwelling) and mass streamfunctions (contour lines, in units of 10$^{10}$ kg\,s$^{-1}$) 
    in the pure N$_2$ (a--b) and N$_2$+CO$_2$ (c--d) experiments. Left and right columns are for CAM3 and ExoCAM, respectively. Panels are shown in tidally-locked coordinates. AP is the antistellar point, and SP is the substellar point.}
    \label{fig:Compare_circ_CAM3-ExoCAM}
\end{figure*}

The atmospheric overturning circulation represented by vertical velocities in pressure coordinates and mass streamfunction is plotted in Figure~\ref{fig:Compare_circ_CAM3-ExoCAM}. There is a very weak and shallow large-scale circulation in the experiment of pure N$_2$ (Figure~\ref{fig:Compare_circ_CAM3-ExoCAM}(a)). The mass streamfunction is about $1\times10^{10}$ kg\,s$^{-1}$. However, in the N$_2$+CO$_2$ experiment, there are well-formed ascending motions over the substellar region and subsidence in the rest of the planet (Figure~\ref{fig:Compare_circ_CAM3-ExoCAM}(c)). The large-scale upwelling and downwelling are connected by outflows towards the nightside in the upper atmosphere and backflows to the substellar region in the lower atmosphere, forming a global overturning circulation with a strength of $150\times10^{10}$ kg\,s$^{-1}$. ExoCAM experiments obtain similar results (Figures~\ref{fig:Compare_circ_CAM3-ExoCAM}(b) and (d)). In ExoCAM, the strengths of the global overturning circulation are $3\times10^{10}$ and $160\times10^{10}$ kg\,s$^{-1}$ for the pure N$_2$ and the N$_2$+CO$_2$ experiments, respectively.

\begin{figure*}[ht]
    \centering
    \includegraphics[width=\textwidth]{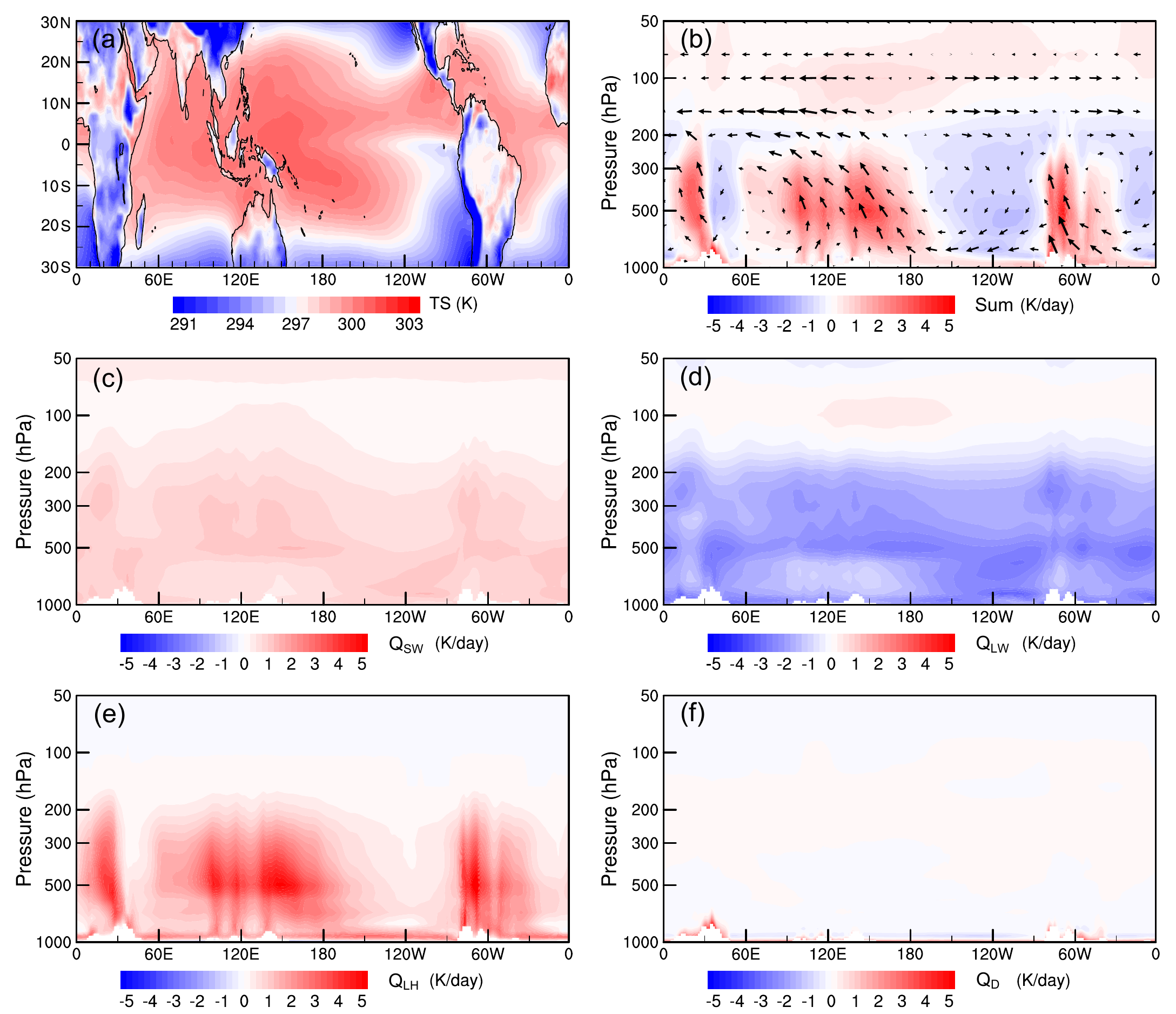}
    \caption{\textcolor{black}{Surface temperature and zonal equatorial atmosphere circulations for modern Earth, simulated by a coupled atmosphere-ocean model CESM \citep{Lanjia_2022}. (a) Tropical surface air temperature. (b) Winds (vectors) and total heating rate (color shading). The vertical components of vectors are magnified by 200 times for clearly showing the overturning structure. (c) Shortwave heating rate, (d) longwave heating rate, (e) heating rate associated with convection process, and (f) turbulent thermal diffusion. All quantities in panels (b--f) are weighted averages between 5$^{\circ}$S and 5$^{\circ}$N.}}
    \label{fig:walker}
\end{figure*}

\textcolor{black}{The global overturning circulation in the N$_2$+CO$_2$ experiment is somewhat stronger than the Walker circulation on Earth. The near-surface wind speed is about 16 m\,s$^{-1}$ in our experiments, while it is about 10 m\,s$^{-1}$ in the tropics of Earth (Figure~\ref{fig:walker}(b)).}

\begin{figure*}
    \centering
    \includegraphics[width=0.6\textwidth]{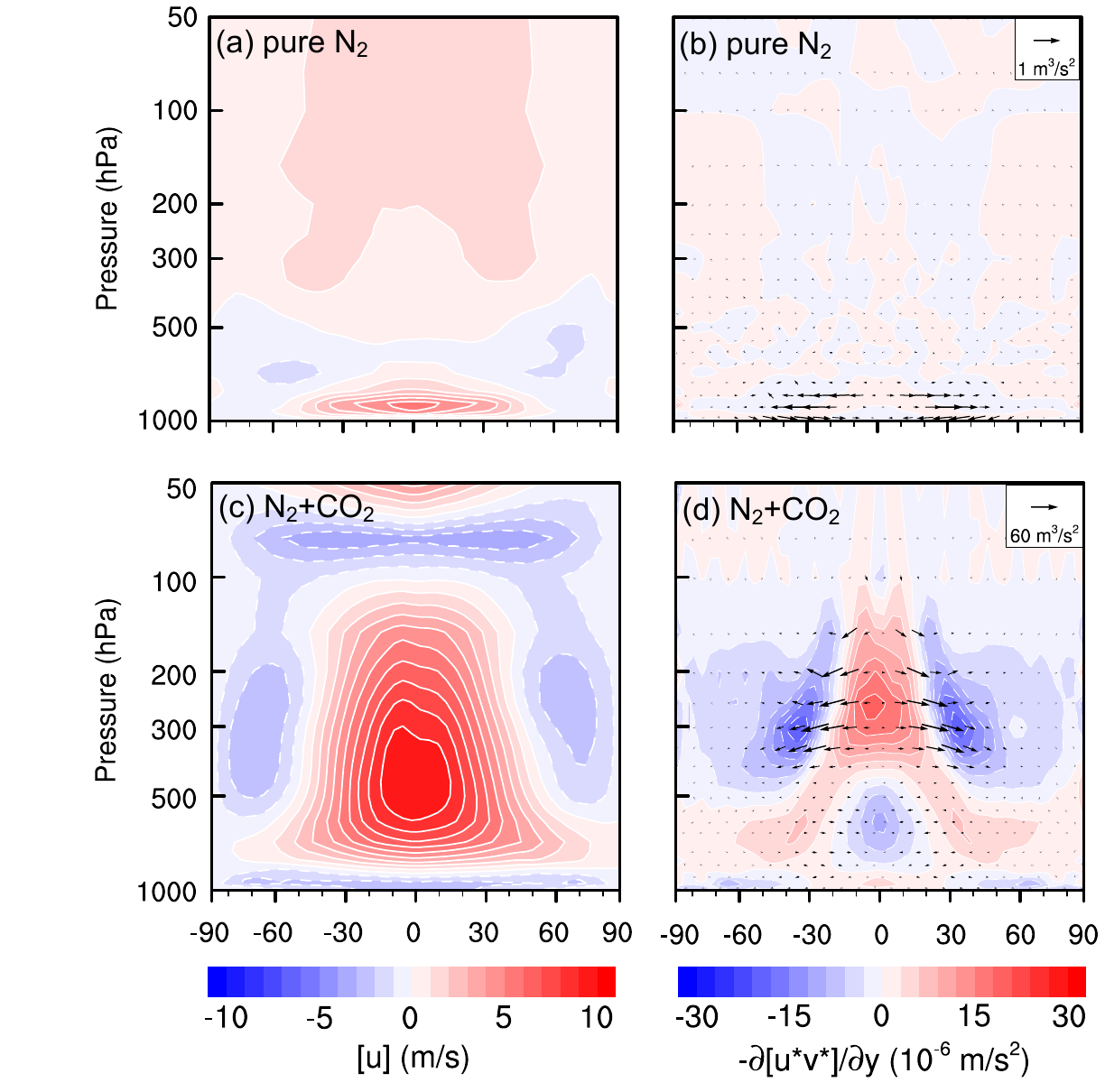}
    \caption{Zonal-mean zonal winds (left) and wave activities (right) in the CAM3 experiments of pure N$_2$ (a--b) and N$_2$+CO$_2$ (c--d). In (b) and (d), color shading is the convergence of eddy momentum flux and vectors are the E-P flux. Note that the arrow scales are different between (b) and (d). These panels are shown in standard coordinates, latitude--pressure.}
    \label{fig:MAIN_circulations}
\end{figure*}

\begin{figure*}
    \centering
    \includegraphics[width=0.8\textwidth]{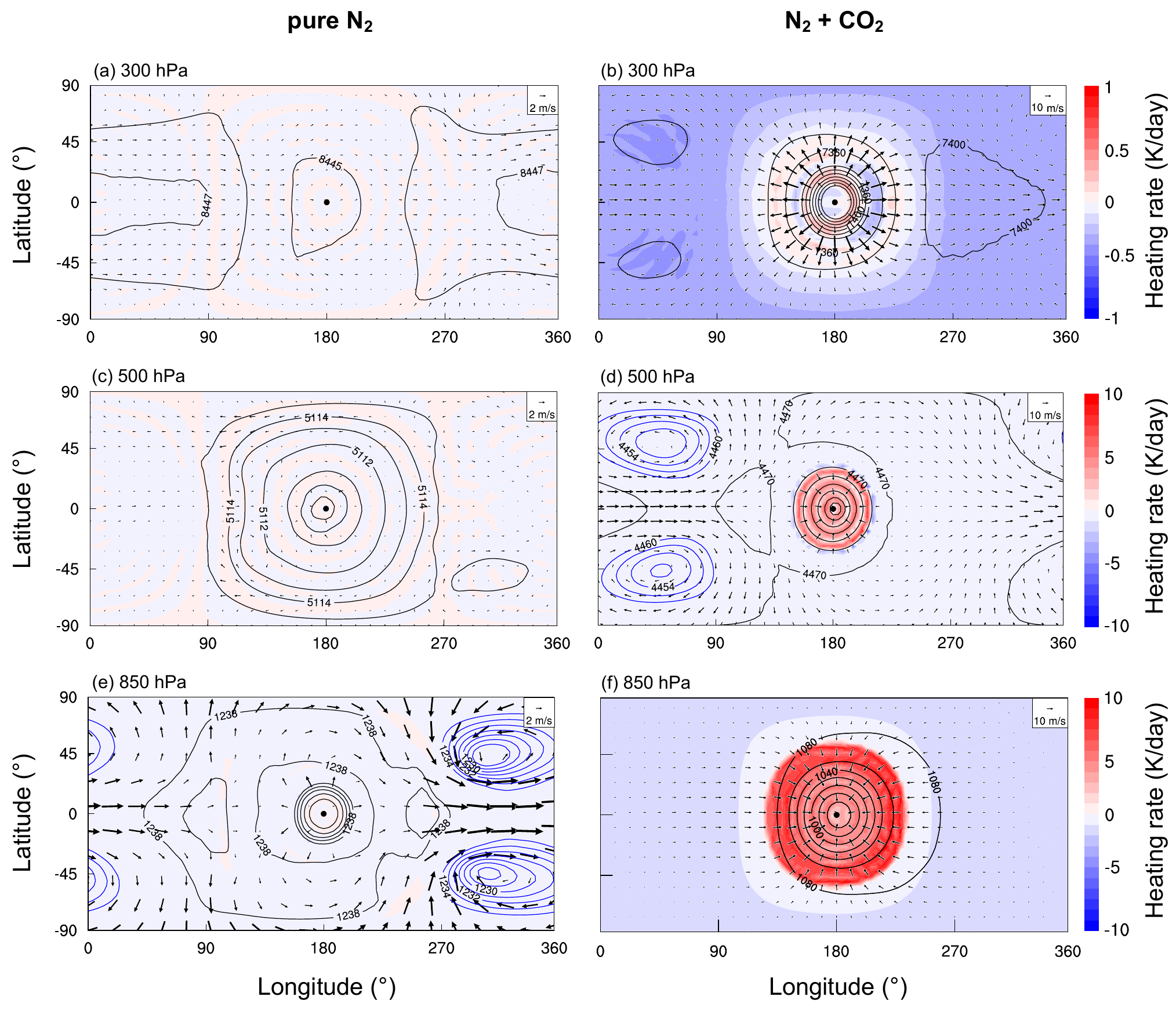}
    \caption{Horizontal wave patterns at 300 hPa (a--b), 500 hPa (c--d), and 850 hPa (e--f) in the CAM3 experiments. Left and right columns are for pure N$_2$ and N$_2$+CO$_2$, respectively. Black dot represents the substellar point. Color shading is the sum of longwave radiative and diffusion heating rates (K/day). Vectors are the horizontal winds (m\,s$^{-1}$). Contour lines are the geopotential height (m), and the intervals are 1 and 20 m for the left and right columns, respectively. The blue contour lines represent the geopotential height in low-pressure Rossby lobes.}
    \label{fig:MAIN_wave_patterns}
\end{figure*}

Another change caused by adding CO$_2$ is the equatorial superrotation, shown in Figure~\ref{fig:MAIN_circulations}. In the pure N$_2$ experiment, the superrotation is only 2--3 m\,s$^{-1}$, and its nominal core is near the surface (Figure~\ref{fig:MAIN_circulations}(a)). While in the N$_2$+CO$_2$ experiment, the maximum jet speed exceeds 10 m\,s$^{-1}$, and the range of the superrotation is from about 850 hPa to the tropopause (about 100 hPa), and one well-formed jet core can be found at about 400 hPa (Figure~\ref{fig:MAIN_circulations}(c)). According to Hide's theorem \citep{Hide_1969}, an up-gradient transport of angular momentum by wave activities is necessary for maintaining the equatorial superrotation against friction. The Eliassen-Palm (E-P) flux is employed to diagnose wave activities \citep{Eliassen_1961,Edmon_1980,Holton_2013}, written as
\begin{equation}\label{eq:E-P_flux}
     \textbf{\emph{F}}=\left[\begin{array}{cc}
        F_y\\
        F_z   
     \end{array}\right]=\left[\begin{array}{cc}
        -\rho\overline{u^*v^*}\\
        \rho f_0R\overline{v^*T^*}/(N^2H)  
     \end{array}\right],
\end{equation}
where $\rho$ is air density, $u$ is zonal wind, $v$ is meridional wind, $T$ is air temperature, $f_0$ is the Coriolis parameter, $R$ is the gas constant, $N$ is buoyancy frequency, and $H$ is the scale height. The stars and overlines represent eddy and zonal-mean components, respectively. The horizontal ($F_y$) and vertical ($F_z$) components of the E-P flux involve the meridional eddy momentum flux and the meridional heat flux, respectively. For convenience, the E-P flux is scaled by a factor of $a\cos\phi/\bar{\rho}$, where $a$ is the planetary radius and $\phi$ is the latitude. The divergence of the E-P flux represents an acceleration of the westerly jet by wave activities \citep{Vallis_2017}. In the pure N$_2$ experiment, the wave activities are weak and located close to the surface (Figure~\ref{fig:MAIN_circulations}(b)). In the N$_2$+CO$_2$ experiment, however, the wave activities are at 100--500 hPa, and their strength substantially enhances, accompanied by an increase in the eastward acceleration (Figure~\ref{fig:MAIN_circulations}(d)). The changes of the superrotation and E-P flux are consistent with the results in \cite{Ding_2019}. They added 0.01-bar CO$_2$ to 1 bar pure O$_2$ atmosphere in their experiments, and showed that the jet speed increases greatly, the position of the jet core moves up, and the E-P flux increases. However, they did not analyze the underlying mechanism.

The E-P flux is mainly contributed from stationary Kelvin and Rossby waves (Figure~\ref{fig:MAIN_wave_patterns}), analogous to a Matsuno-Gill mode \citep{Matsuno_1966,Gill_1980,Showman_2011,Wang_2020}. In the pure N$_2$ experiment, the layer of 850 hPa clearly shows stationary waves (Figure~\ref{fig:MAIN_wave_patterns}(e)). However, the stationary waves disappear in other layers such as 300 and 500 hPa (Figures~\ref{fig:MAIN_wave_patterns}(a) and (c)). The E-P flux therefore only exists in the very low-level layers (Figure~\ref{fig:MAIN_circulations}(b)). In the N$_2$+CO$_2$ experiment, the stationary waves are obvious throughout the free troposphere (Figures~\ref{fig:MAIN_wave_patterns}(b) and (d)). Moreover, the stationary waves are much stronger. For example, the amplitude of the waves in geopotential height at 500 hPa is about 120 m in the N$_2$+CO$_2$ experiment (Figure~\ref{fig:MAIN_wave_patterns}(d)), while the value in the pure N$_2$ experiment is only about 7 m (Figure~\ref{fig:MAIN_wave_patterns}(c)). The larger wave amplitude can explain the stronger E-P flux and the greater acceleration on the jet. Figure~\ref{fig:MAIN_wave_patterns} also shows that the horizontal divergence/convergence in the N$_2$+CO$_2$ experiment is much stronger than in the pure N$_2$ experiment.

\subsection{\textcolor{black}{Mechanism for the atmospheric overturning circulation}}\label{sec:results_3}

\begin{figure*}[ht]
    \centering
    \includegraphics[width=0.9\textwidth]{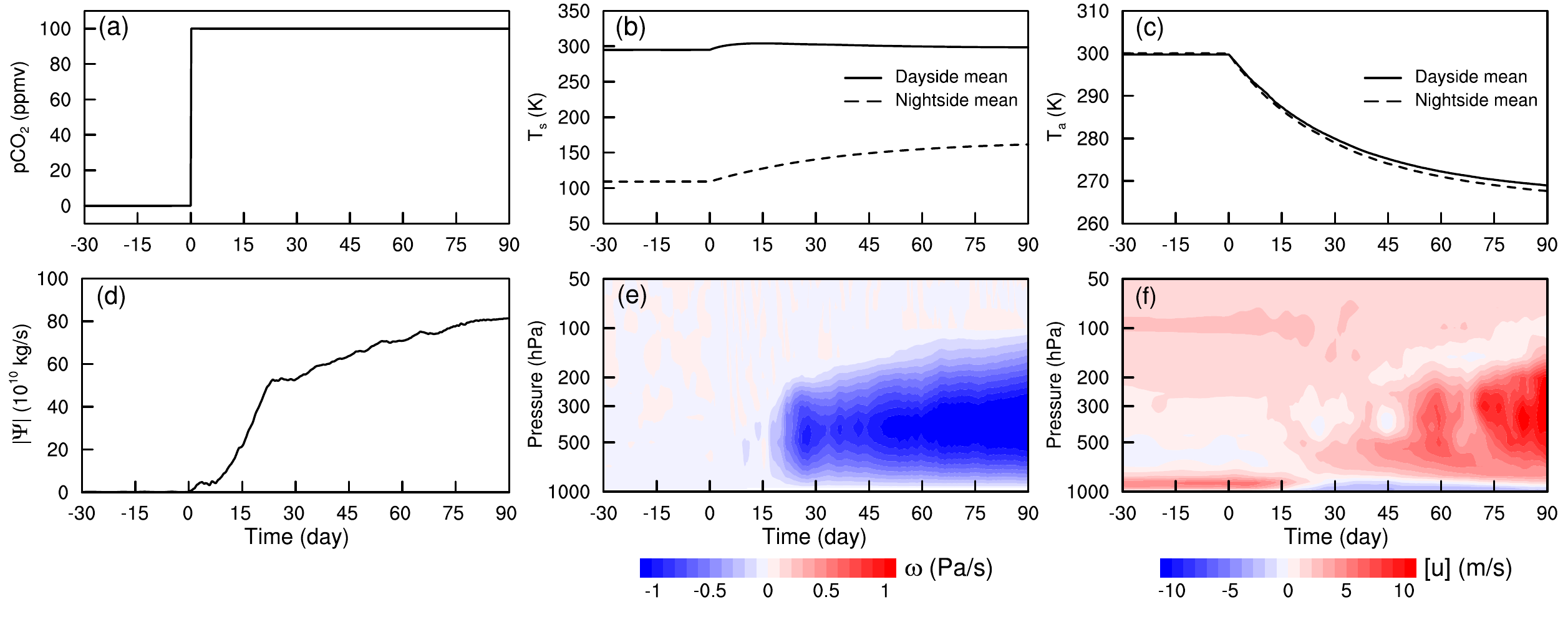}
    \caption{The abrupt transition experiment from a pure N$_2$ atmosphere to a N$_2$+CO$_2$ atmosphere. Time series of (a) CO$_2$ concentration, (b) surface temperature, (c) vertically-integrated mass-weighted air temperature, (d) the strength of mass streamfunction, (e) vertical velocity in pressure coordinates at the substellar point, and (f) zonal-mean zonal wind at the equator. Solid and dashed lines in (b) and (c) represent the dayside-mean and nightside-mean, respectively. CO$_2$ of 100 ppmv is added to the atmosphere on Day 0.}
    \label{fig:MAIN_branch-run}
\end{figure*}

What mechanism drives the strong atmospheric overturning circulation in the N$_2$+CO$_2$ experiment? Meanwhile, why is the overturning circulation so weak in the pure N$_2$ experiment? To answer these questions, we firstly design a branch experiment to further examine whether it is CO$_2$ that drives the overturning circulation. We suddenly add 100-ppmv CO$_2$ to the equilibrium state of the pure N$_2$ experiment (Figure~\ref{fig:MAIN_branch-run}(a)). After adding CO$_2$, the global overturning circulation appears and becomes stronger as a function of time (Figure~\ref{fig:MAIN_branch-run}(d)). On the 15$^{\rm{th}}$ model day, an obvious upwelling over the substellar region is formed (Figure~\ref{fig:MAIN_branch-run}(e)). The equatorial superrotation also begins to appear at this time (Figure~\ref{fig:MAIN_branch-run}(f)). Within about 90 model days, the global overturning circulation and the equatorial superrotation are well established. This circulation transition from the pure N$_2$ scenario to the N$_2$+CO$_2$ scenario confirms that the driving mechanism is associated with CO$_2$.

\begin{figure*}[ht]
    \centering
    \includegraphics[width=0.9\textwidth]{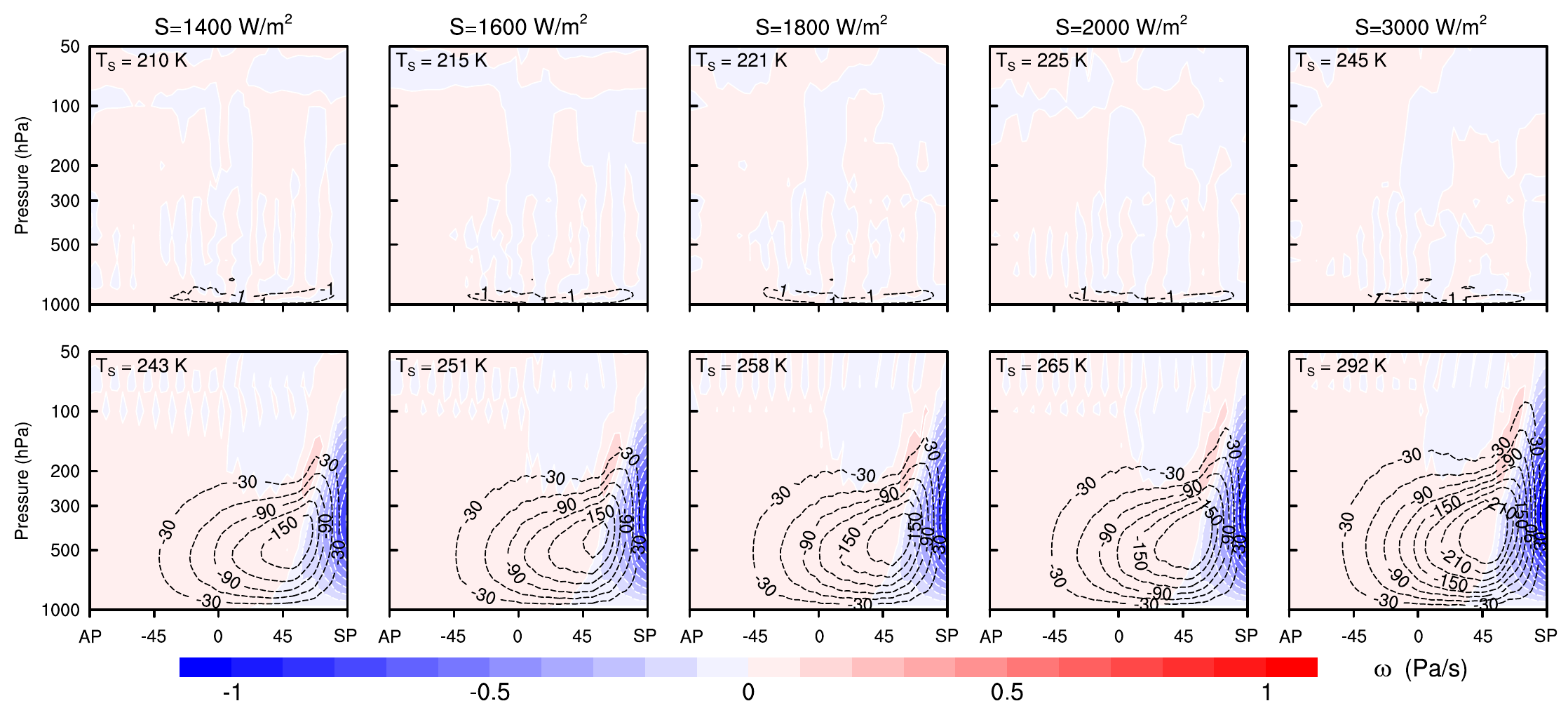}
    \caption{Vertical velocities in pressure coordinates (color shading, negative: upwelling, and positive: downwelling) and mass streamfunctions (contour lines, in units of 10$^{10}$ kg\,s$^{-1}$) for pure N$_2$ (upper row) and N$_2$+CO$_2$ (lower row). The stellar fluxes at the substellar point are 1400, 1600, 1800, 2000, and 3000 W\,m$^{-2}$ (from left to right). The global-mean surface temperature of each simulation is shown at the top-left corner. The coordinates are same as those in Figure~\ref{fig:Compare_circ_CAM3-ExoCAM}.}
    \label{fig:MAIN_zmpsi_diffS}
\end{figure*}

The existence of CO$_2$ makes the surface warmer (greenhouse effect) and meanwhile the atmosphere colder (radiative cooling effect), as shown in Figures~\ref{fig:MAIN_branch-run}(b \& c),~\ref{fig:MAIN_surf-vert_temp} \& \ref{fig:MAIN_ExoCAM} and Section~\ref{sec:results_1}. Which one of them is the key driving mechanism for the atmospheric circulation? In order to answer this question, we design a series of experiments within which the stellar flux is increased from 1200 to 1400, 1600, 1800, 2000, or 3000 W\,m $^{-2}$, to make the surface warmer. Two scenarios, pure N$_2$ and N$_2$+CO$_2$, are run for each stellar flux. In the pure N$_2$ experiments, the increased stellar flux warms the surface, but the global overturning circulation is still very weak (upper row in Figure~\ref{fig:MAIN_zmpsi_diffS}). In the N$_2$+CO$_2$ experiments, a strong global overturning circulation exists under all stellar fluxes (bottom row in Figure~\ref{fig:MAIN_zmpsi_diffS}). These experiments indicate that a warmer surface cannot directly drive the global overturning circulation. They also uncover that the uneven distribution of stellar energy is not the direct process driving the atmospheric circulation, an intermediate `medium' is required.

\begin{figure*}[ht]
    \centering
    \includegraphics[width=0.9\textwidth]{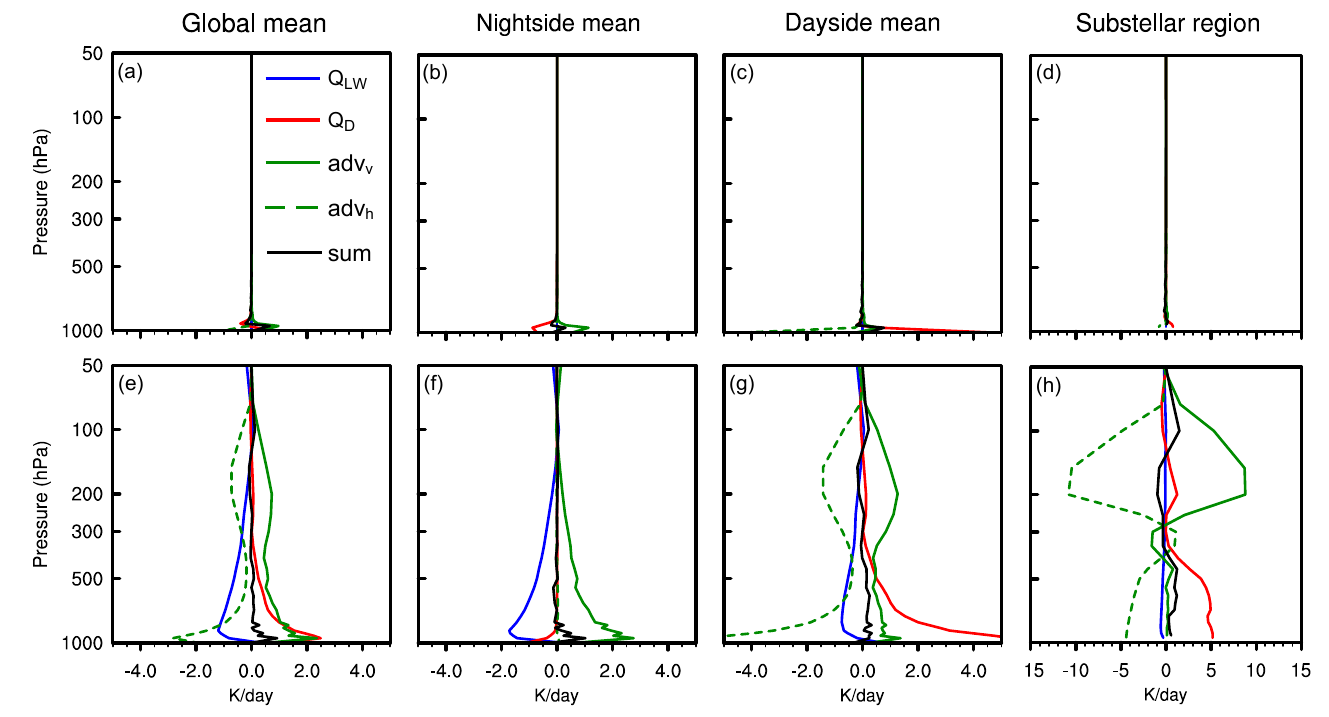}
    \caption{Vertical profiles of heating rates in the CAM3 experiments of pure N$_2$ (a--d) and N$_2$+CO$_2$ (e--h). (a \& e) global-mean; (b \& f) nightside-mean; (c \& g) dayside-mean; (d \& h) substellar region. Blue: heating rate from longwave radiation ($Q_{LW}$); red: turbulent thermal diffusion ($Q_D$); solid green: vertical advection ($adv_v$); dashed green: horizontal advection ($adv_h$); and black: the sum.}
    \label{fig:MAIN_heating-rate}
\end{figure*}

In order to know the underlying intermediate `medium', atmospheric thermal equation in pressure coordinates is analyzed, written as
\begin{equation}\label{eq:T-tendency}
    \frac{\partial T}{\partial t}+\left(u\frac{\partial}{\partial x}+v\frac{\partial}{\partial y}\right)T-S_p\omega=Q,
\end{equation}
where $Q$ is the diabatic heating rate, $S_p=RT/(pc_p)-\partial T/\partial p$ is the static stability ($c_p$ is the specific heat at constant pressure, and $p$ is the pressure), $T$ is the air temperature, and $\omega$ is the vertical velocity in pressure coordinates. The second and third terms on the left side are horizontal ($adv_h$) and vertical thermal advections ($adv_v$), respectively. This equation is similar to Equation (2.43) in \cite{Holton_2013} and Equation (4) in \cite{Fujii_2017}. The absorption of the shortwave radiation has a negligible contribution in the experiments since there is no water vapor or other shortwave absorbers (CO$_2$ molecule can absorb somewhat near-infrared energy, but its strength is too weak), and latent heat is absent on completely dry planets. The diabatic heating rate ($Q$) in Equation~(\ref{eq:T-tendency}) includes the contributions of longwave radiation ($Q_{LW}$) and turbulent thermal diffusion ($Q_{D}$). Note that the turbulent thermal diffusion is from boundary layer parameterization scheme, which describes the diffusive effects of turbulence and sub-scale dry convection \citep{Byron_2003}.

Figure~\ref{fig:MAIN_heating-rate} shows the contribution of each term in Equation~(\ref{eq:T-tendency}) for a steady state. The total heating rate (black lines) is close to zero; this suggests that the experiments have reached equilibrium. In the pure N$_2$ experiment, the longwave cooling is negligible (blue lines), because the atmosphere is transparent to longwave radiation (Figures~\ref{fig:MAIN_heating-rate}(a--d)). The turbulent thermal diffusion (red lines) and the thermal advection (green lines) are also very small. In contrast, there is a strong radiative cooling in the N$_2$+CO$_2$ experiment (Figure~\ref{fig:MAIN_heating-rate}(e)). In global mean, it is balanced by the turbulent thermal diffusion. Moreover, Figure~\ref{fig:MAIN_heating-rate}(f) shows that the radiative cooling is dominant on the nightside. Note that the horizontal range of radiative-cooling is much larger than the night hemisphere, because the temperature inversion covers  both nightside and most region of the dayside except the substellar region (Figures~\ref{fig:MAIN_surf-vert_temp} and \ref{fig:MAIN_ExoCAM}). Over the substellar region, the turbulent thermal diffusion is dominated (Figure~\ref{fig:MAIN_heating-rate}(h)).

\begin{figure*}[ht]
    \centering
    \includegraphics[width=\textwidth]{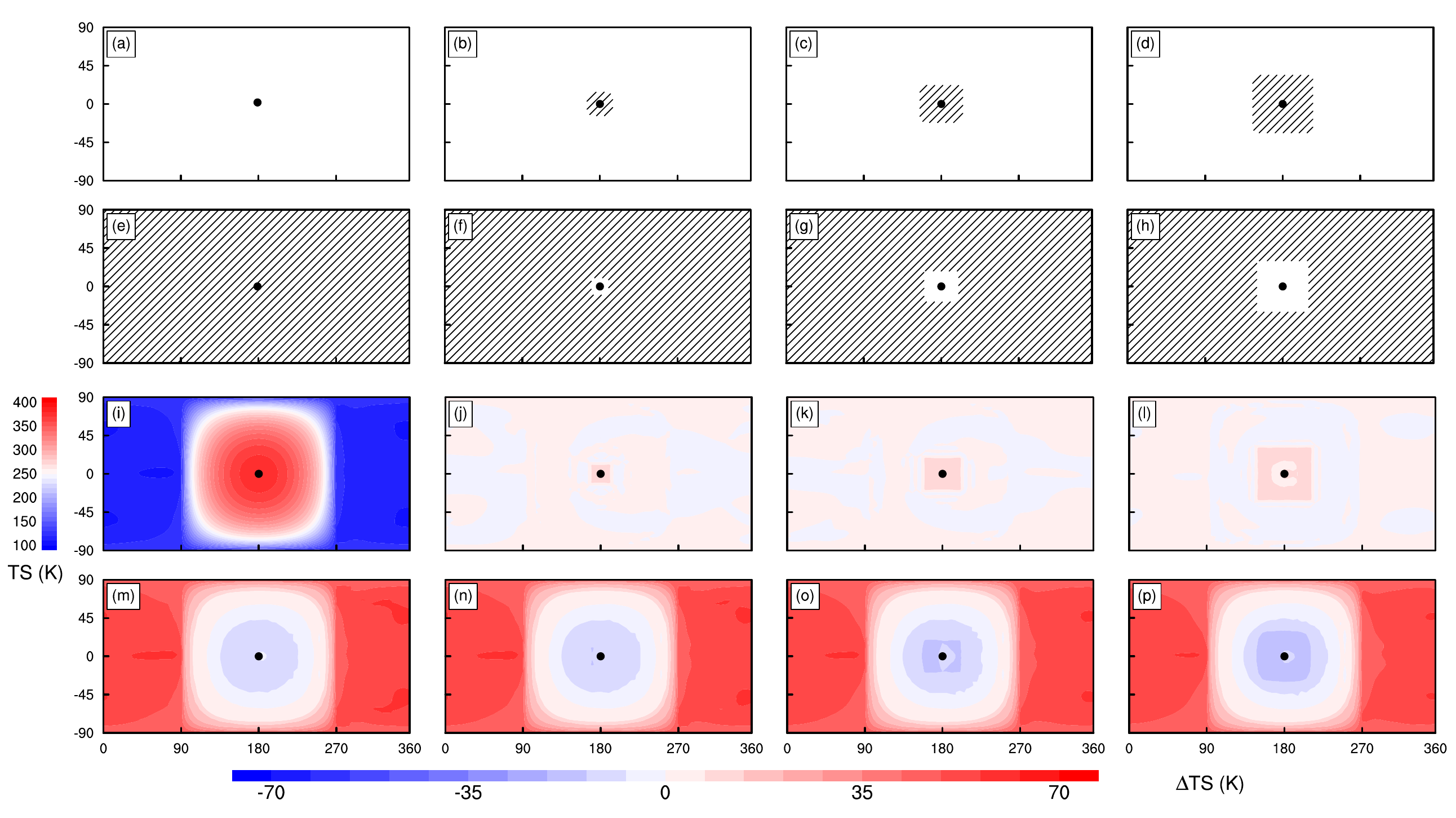}
    \caption{Spatial distributions of CO$_2$ concentration (a--h) and corresponding surface temperature (i--p) in regionally-prescribed CO$_2$ experiments. CO$_2$ concentration is 100 ppmv in the region filled with oblique lines and zero in the white region. The ranges of filled regions in (b--d) are 10$^{\circ}$, 20$^{\circ}$, and 30$^{\circ}$ of longitudes and latitudes around the substellar point. In opposite, the non-filled regions in (f--h) are 10$^{\circ}$, 20$^{\circ}$, and 30$^{\circ}$ of longitudes and latitudes around the substellar point. Panel (i) is the surface temperature without any CO$_2$, and panels (j--p) are the differences of the surface temperature from that in panel (i) for the experiments of (b--h), respectively. The four panels of the leftmost row is the same as the control experiments shown in section~\ref{sec:results_1}.}
    \label{fig:prescribed_co2_TS}
\end{figure*}

\begin{figure*}[ht]
    \centering
    \includegraphics[width=0.9\textwidth]{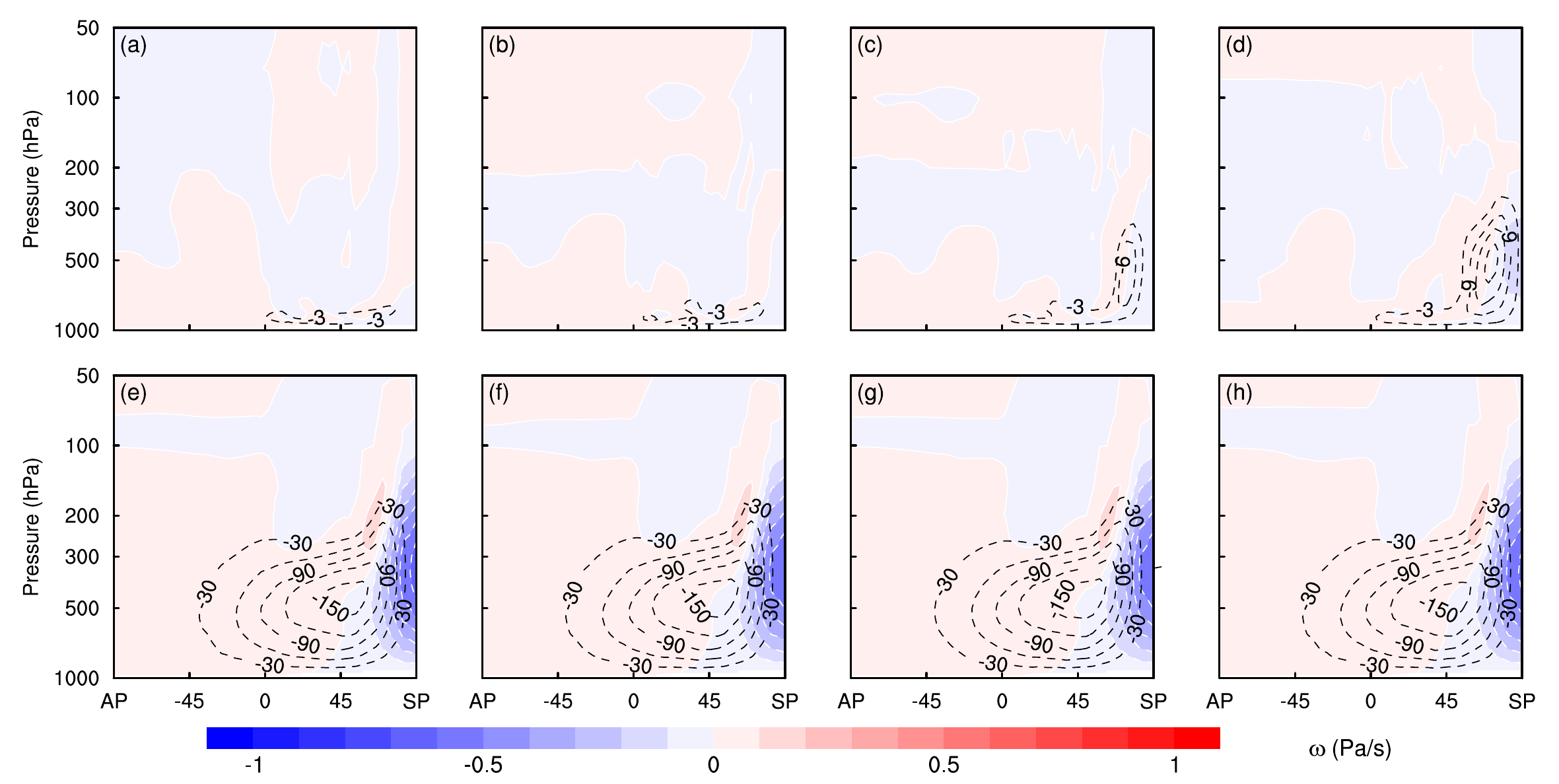}
    \caption{Vertical velocities in pressure coordinates (color shading) and mass streamfunctions (contour lines, 10$^{10}$ kg\,s$^{-1}$) in regionally-prescribed CO$_2$ experiments. Distributions of CO$_2$ concentration for (a--h) correspond to those in Figures~\ref{fig:prescribed_co2_TS}(a--h), respectively. The coordinates are same as those in Figure~\ref{fig:Compare_circ_CAM3-ExoCAM}.}
    \label{fig:prescribed_co2_circ}
\end{figure*}

\begin{figure*}[ht]
    \centering
    \includegraphics[width=0.87\textwidth]{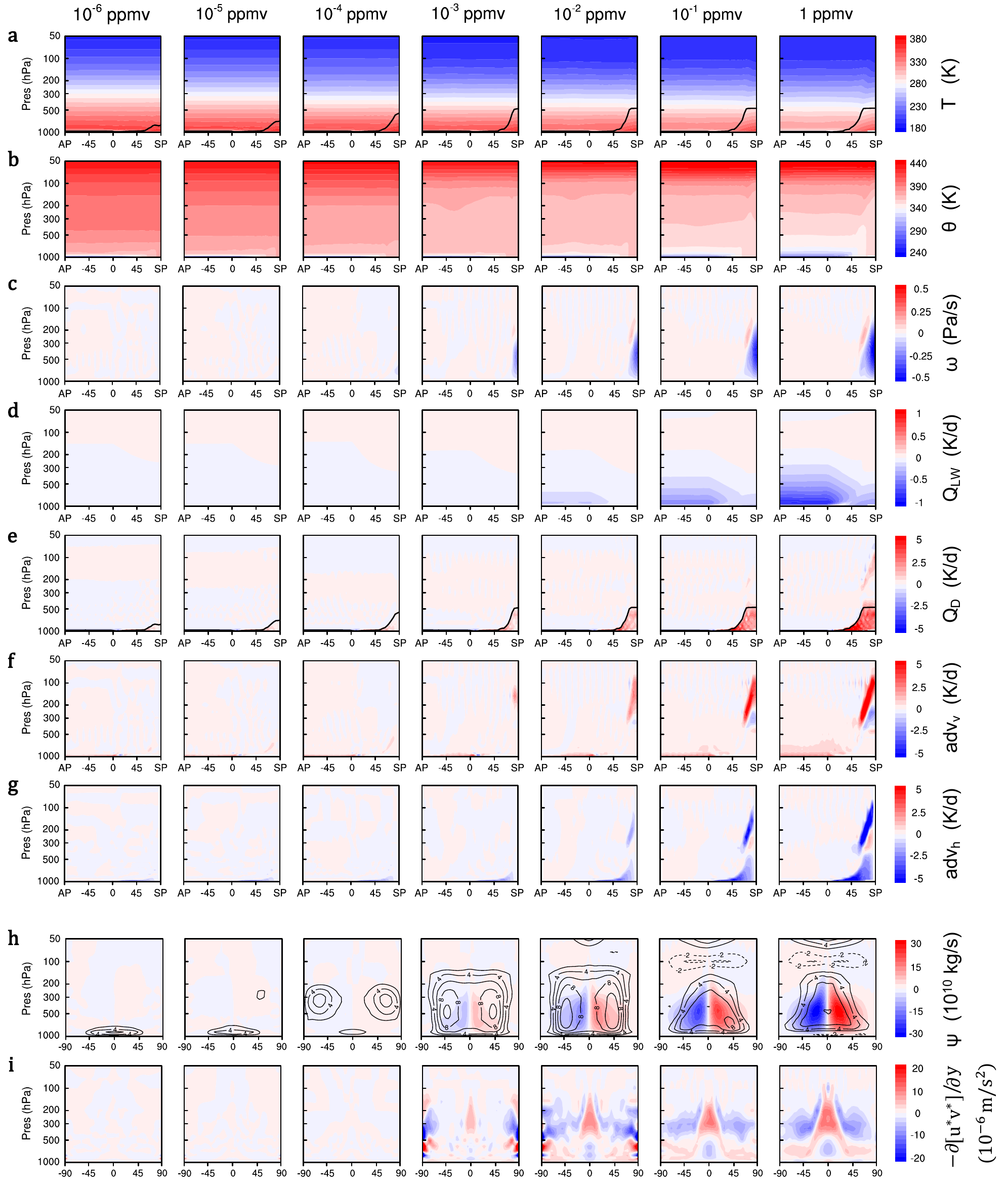}
    \caption{(a) Temperature, (b) potential temperature, (c) vertical velocity in pressure coordinates, (d) heating rate from longwave radiation, (e) turbulent thermal diffusion, (f) vertical thermal advection, (g) horizontal thermal advection, (h) mass streamfunction (color shading) and zonal-mean zonal winds (contour lines, interval is 2 m\,s$^{-1}$), and (i) the acceleration on the zonal winds caused by wave activities. CO$_2$ concentrations are 10$^{-6}$, 10$^{-5}$, 10$^{-4}$, 10$^{-3}$, 10$^{-2}$, 10$^{-1}$, and 1 ppmv (from left to right). Black line in (a) and (e) represents the height of the planetary boundary layer. Panels (a) to (g) are shown in tidally-locked coordinates and are averaged over tidally-locked longitude, AP is the antistellar point, and SP is the substellar point; panels (h) and (i) are in standard coordinates.}
    \label{fig:MAIN_climate-diff-CO2}
\end{figure*}

Local diabatic heating and radiative cooling drive the global overturning circulation. In the cooling region, the vertical velocity in pressure coordinates ($\omega$) is approximately equal to $-Q/S_p$, and $\omega$ is larger than zero (downwelling) because $Q$ is dominated by $Q_{LW}<0$. This indicates that the longwave cooling drives the robust large-scale downwelling. Meanwhile, the downwelling could force upwelling around the substellar point according to mass conservation. In the substellar region, the turbulent thermal diffusion warms the atmosphere and has a partial contribution to the upwelling there.

\textcolor{black}{For the Walker circulation on Earth, the cooling of equatorial atmosphere is also mainly due to the longwave emission (Figure~\ref{fig:walker}(d)), while shortwave absorption by water vapor (Figure~\ref{fig:walker}(c)) and latent heat release (Figure~\ref{fig:walker}(e)) act to warm the equatorial atmosphere. The net pattern is that there is warming in the upwelling branches and cooling in the downwelling branches (Figure~\ref{fig:walker}(b)). This is consistent with the driving mechanism mentioned above, except that the shortwave absorption and latent heat release are nearly zero on dry rocky planets.}

In order to quantify the respective contributions of radiative cooling and turbulent thermal diffusion to the global overturning circulation, we modify the model CAM4 to run with regionally-prescribed CO$_2$. It means that over one region of the global the atmosphere is prescribed with non-zero CO$_2$ and in the rest region the CO$_2$ concentration is set to zero. Six experiments were run. Three of them have 100-ppmv CO$_2$ in the region of respectively 10$^{\circ}$, 20$^{\circ}$, and 30$^{\circ}$ of longitudes and latitudes around the substellar point and the rest regions have zero CO$_2$ (Figures~\ref{fig:prescribed_co2_TS}(b--d)). The other three have the opposite setup, i.e., in the region of respectively 10$^{\circ}$, 20$^{\circ}$, and 30$^{\circ}$ of longitudes and latitudes around the substellar point the CO$_2$ concentration is zero, and the rest regions have 100-ppmv CO$_2$ (Figures~\ref{fig:prescribed_co2_TS}(f--h)).

Results show that the regionally-prescribed CO$_2$ indeed warms the surface of the corresponding regions via greenhouse effect (Figures~\ref{fig:prescribed_co2_TS}(j--l \& n--p)). When 100-ppmv CO$_2$ is prescribed in the substellar region, the overturning circulation is narrow and weak, with a strength of $\approx$$3\times10^{10}$ to $12\times10^{10}$ kg\,s$^{-1}$ (Figures~\ref{fig:prescribed_co2_circ}(b--d)). This suggests that the contribution of turbulent thermal diffusion to the overturning circulation is small, although not zero. Whereas, in the experiments with prescribed CO$_2$ outside of the substellar region, there is a strong global overturning circulation, with a strength of $\approx$$150\times10^{10}$ kg\,s$^{-1}$. This suggests that the contribution of radiative cooling to the overturning circulation is significant (Figures~\ref{fig:prescribed_co2_circ}(f--h)). In combination, these six experiments suggest that the global overturning circulation is mainly driven by CO$_2$ radiative cooling outside of the substellar region.

A question is how much CO$_2$ is required to drive a robust global overturning circulation? In order to answer this question, we test a series of CO$_2$ concentration, 10$^{-6}$, 10$^{-5}$, 10$^{-4}$, 10$^{-3}$, 10$^{-2}$, 10$^{-1}$, and 1 ppmv. As the CO$_2$ concentration is increased, the radiative cooling effect becomes stronger (Figure~\ref{fig:MAIN_climate-diff-CO2}(d)), causing the atmospheric temperature and potential temperature to decrease (Figures~\ref{fig:MAIN_climate-diff-CO2}(a \& b)), although the surface temperature increases. Larger longwave cooling drives stronger global overturning circulation (Figure~\ref{fig:MAIN_climate-diff-CO2}(c \& d)). The turbulent thermal diffusion and advections also become stronger (Figures~\ref{fig:MAIN_climate-diff-CO2}(e--g)).

The required CO$_2$ concentration driving a robust global overturning circulation is small, between 10$^{-3}$ and 10$^{-2}$ ppmv (see Figures~\ref{fig:MAIN_climate-diff-CO2}(c), (h), and (i)). CO$_2$ concentrations in previous simulations of dry, tidally locked rocky planets, such as a pure CO$_2$ atmosphere in \cite{Joshi_1997}, 345 ppmv in \cite{Edson_2011}, 376 ppmv in \cite{Leconte_2013}, pure CO$_2$ in \cite{Wordsworth_2015}, and 10$^4$ ppmv in \cite{Ding_2019}, are much larger than the required limit, so strong global overturning circulation and equatorial superrotation exist in all their experiments.

Note that Figure~\ref{fig:MAIN_climate-diff-CO2}(h) shows the state transition from double jets to one well-formed superrotation when CO$_2$ concentration is increased from 10$^{-3}$ to 1 ppmv. This phenomenon is caused by the competition between the poleward angular momentum transport by large-scale meridional circulation and the equatorward momentum transport by tropical wave activities. As CO$_2$ concentration is increased to 10$^{-3}$ ppmv, the poleward angular momentum transport becomes effective associated with the formation of the meridional circulation. In cases with 10$^{-3}$- and 10$^{-2}$-ppmv CO$_2$, the tropical wave activities are relatively weak, so that the poleward angular momentum transport dominates, maintaining one mid-latitude jet on each hemisphere. When CO$_2$ concentration is over 10$^{-1}$ ppmv, the tropical wave activities become stronger and dominate (Figure~\ref{fig:MAIN_climate-diff-CO2}(i)), so that a well-formed equatorial superrotation appears (Figure~\ref{fig:MAIN_climate-diff-CO2}(h)).

\subsection{Rapidly rotating planets}\label{sec:results_4}

\begin{figure*}[ht]
    \centering
    \includegraphics[width=0.7\textwidth]{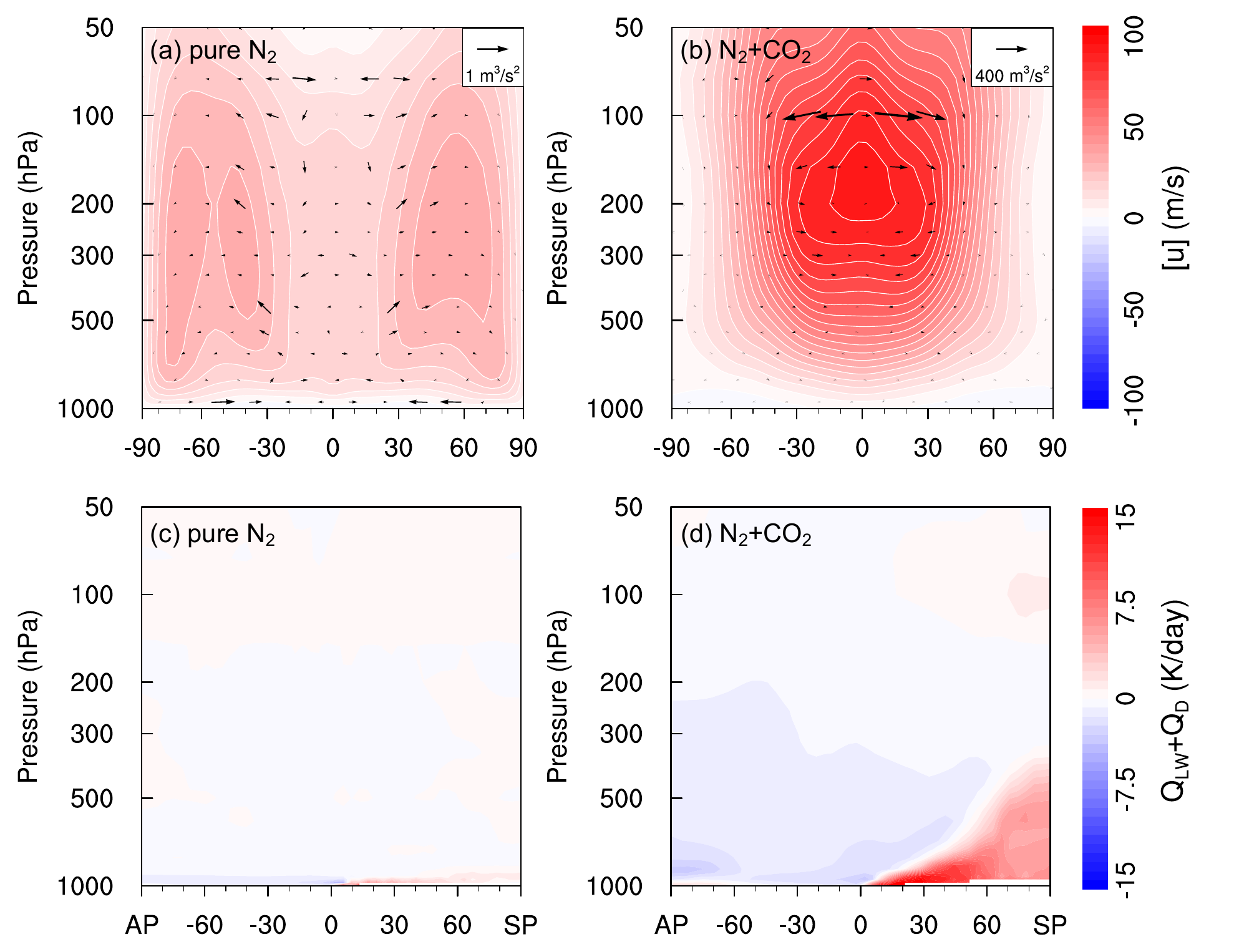}
    \caption{Results for planets with pure N$_2$ atmosphere (left) and N$_2$+CO$_2$ atmosphere (right) under a rotation period of 5 Earth days. (a) and (b): zonal-mean zonal winds and the scaled E-P flux, same as those in Figure~\ref{fig:MAIN_circulations}. (c) and (d): the sum of heating rates from longwave radiation and turbulent thermal diffusion averaged over tidally-locked longtitude. AP is the antistellar point, and SP is the substellar point.}
    \label{fig:MAIN_climate-fast-period}
\end{figure*}

Rapidly rotating tidally-locked planets are in a different dynamical regime from the slowly rotating planets \citep{Merlis_2010,Edson_2011,Carone_2014,Carone_2015,Carone_2016,Noda_2017,Haqq_Misra_2018}. In order to examine whether the driving mechanism on slowly rotating planets is applicable to rapidly rotating planets, we change the rotation period to 5 Earth days, and re-run the pure N$_2$ and N$_2$+CO$_2$ experiments. In these two experiments, the circulation is a combination of a global overturning circulation and a zonally symmetric circulation \citep{Merlis_2010,Noda_2017,Haqq_Misra_2018,Hammond_2021}. Due to this reason, mass streamfunction is no longer an appropriate representation of the atmospheric circulation, so that below we focus on the jet streams, as shown in Figure~\ref{fig:MAIN_climate-fast-period}.

In the pure N$_2$ experiment, there are two mid-latitude jets with wind speed of about 30 m\,s$^{-1}$, and a relatively weak equatorial jet with wind speed of about 15 m\,s$^{-1}$ (Figure~\ref{fig:MAIN_climate-fast-period}(a)). This is because the rapid rotation makes the Rossby deformation radius be smaller than half of the planetary radius, and then the planet is in a mixed dynamical state controlled by both tropical and extratropical Rossby waves, while the latter is stronger. The horizontal wave pattern in the pure N$_2$ experiment (figure not shown) well matches the typical pattern of mixed tropical and extratropical Rossby waves that is shown in Figure 12 of  \cite{Carone_2015}. The shear between the tropical Rossby and Kelvin waves maintains the weak equatorial superrotation. Meanwhile, the momentum transports by the large-scale meridional circulation and by the coupled extratropical Rossby and tropical Kelvin waves maintain the mid-latitide jets. In the N$_2$+CO$_2$ experiment, there is a very strong superrotation with one well-formed jet core instead of two mid-latitude jets, and the speed of the equatorial superrotation exceeds 100 m\,s$^{-1}$ (Figure~\ref{fig:MAIN_climate-fast-period}(b)). This is mainly related to the dramatically enhanced longwave cooling in the atmosphere (comparing Figures~\ref{fig:MAIN_climate-fast-period}(c) and (d)). It drives intense upwelling over the substellar region and strong tropical waves. The horizontal wave patterns are dominated by equatorial Rossby and Kelvin waves, as well as the equatorward momentum transport (figure not shown). As a consequence, there is a large divergence of E-P flux above the equator which maintains the superrotation (vectors in Figure~\ref{fig:MAIN_climate-fast-period}(b)). These two experiments demonstrate that the driving mechanism on slowly rotating planets is applicable to rapidly rotating planets although the circulation is more complex.

\section{Summary}\label{sec:summary}
\begin{figure*}[ht]
    \centering
    \includegraphics[width=\textwidth]{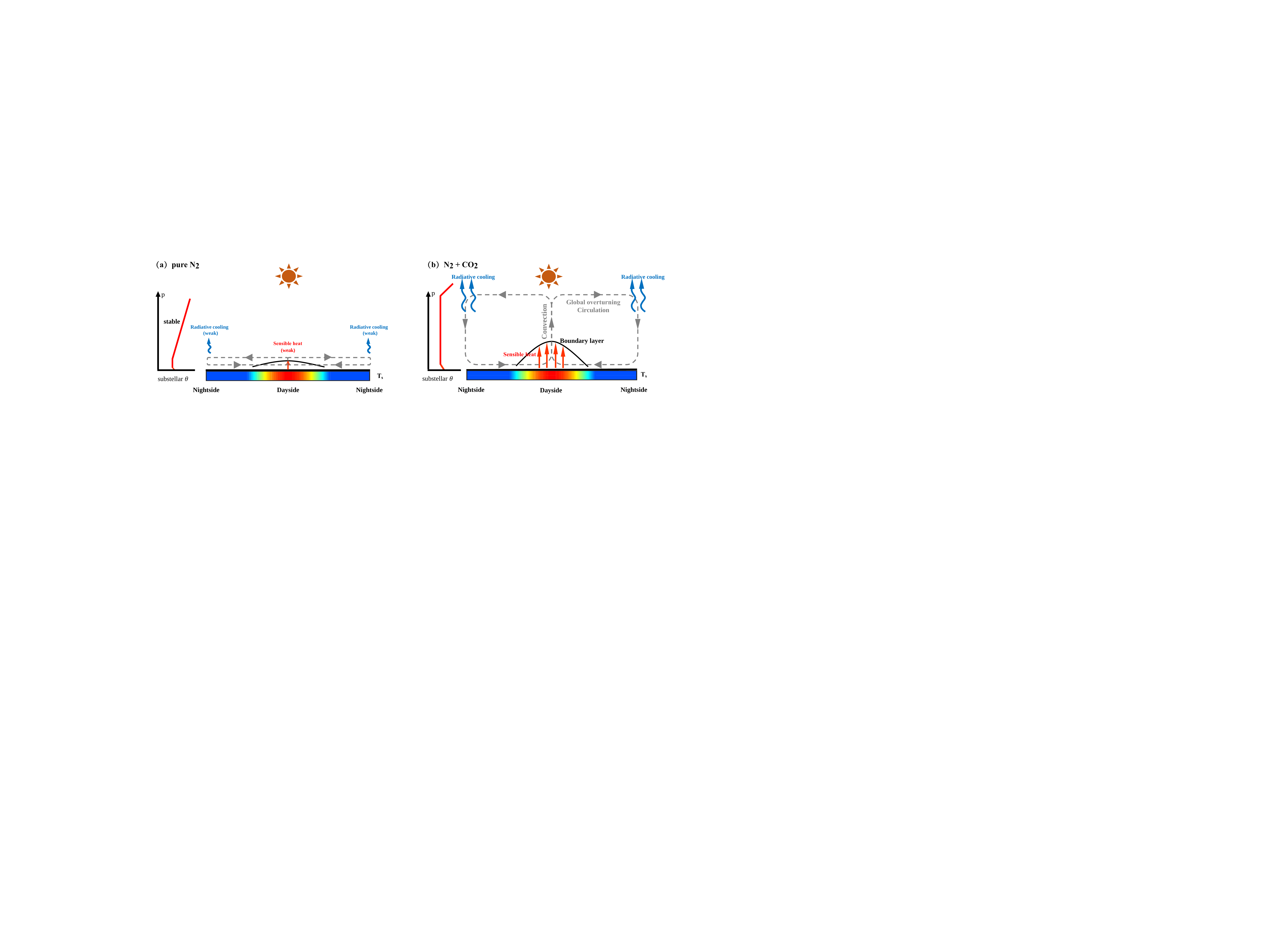}
    \caption{Atmospheric overturning circulation diagram, same as Figure~\ref{fig:clima_diag} but for dry, slowly rotating, tidally-locked rocky planets with atmosphere of pure N$_2$ (a) and N$_2$+CO$_2$ (b).}
    \label{fig:sum_diagram}
\end{figure*}

In this study, the atmospheric overtruning circulation on dry, slowly-rotating, tidally-locked terrestrial planets is investigated through numerical simulations and theoretical analyses. Three models are employed, including CAM3, CAM4, and ExoCAM. A summary of the main findings is shown in Figure~\ref{fig:sum_diagram}. The atmospheric overturning circulation on dry, tidally locked terrestrial planets is mainly driven by CO$_2$ (or equivalently other greenhouse gas(es)) radiative cooling over the regions of temperature inversion, rather than by radiative warming over the substellar region. The uneven distribution of stellar radiation is the ultimate mechanism but not the direct triggering mechanism. If there is no greenhouse gas, the overturning circulation would be extremely weak. Stellar radiation cannot be the direct triggering mechanism; this is because the dry atmosphere is nearly transparent to stellar radiation, as long as O$_3$ and O$_2$ are not considered. The required concentration of CO$_2$ to maintain a strong overturning circulation is small, about 10$^{-2}$~ppmv. This implies that most of dry, tidally locked rocky planets should have a strong overturning circulation with upwelling over the substellar region and downwelling in the rest of the global. On dry, rapidly rotating, tidally-locked terrestrial planets, CO$_2$ radiative cooling is also important for the atmospheric circulation, although baroclinic instability in mid-latitudes becomes another important factor.

This mechanism discussed in this study is applicable to Venus, because there is strong net radiative cooling induced by CO$_2$ at high latitudes of Venus \citep{Crisp_1989,Haus_2017,Limaye_2018}. Besides radiative cooling, radiative heating induced by sunlight absorption can also drive atmospheric circulations on dry planets. For instance, absorption by the ultraviolet absorbers and sulphuric acid cloud particles creates net radiative heating at low latitudes, which avails to drive upwelling and the Venusian Hadley circulation \citep{Tomasko_1985,Crisp_1989,Limaye_2018}. For Mars, during its summer and winter seasons, lofted dust absorbs solar radiation and heats the atmosphere in the summer hemisphere, which drives a solstitial Hadley circulation extending from the summer pole to the winter pole \citep{Wilson_1997,Forget_1999,Newman_2005,Hasegawa_2012,climate_of_mars}.

In our experiments, the global overturning circulation becomes stronger as CO$_2$ concentration is increased (see Figure~\ref{fig:MAIN_climate-diff-CO2}), because its  radiative cooling enhances. Interestingly, an opposite trend is for the Hadley circulation on Earth, which tends to become weaker as CO$_2$ concentration is increased under global warming  \citep[e.g.,][]{ipcc2021}. On Earth, water vapor feedback is important, which acts to reduce tropical lapse rate and increase the vertical stratification, weakening the Hadley circulation \citep[e.g.,][]{Held_2006,Ma_2012,Bony_2013,Vallis_Zurita_2015,Chemke_2021}.

In this work, we have run pure N$_2$ and N$_2$+CO$_2$ experiments with different stellar fluxes and rotation periods. These experiments indicate that radiative cooling is an efficient mechanism for driving the atmospheric overturning circulation on dry, tidally-locked rocky planets. While, other factors, such as the spectrum of host star, planetary radius, gravity, atmospheric molecular weight, and surface pressure, may also important \citep{Leconte_2013,Carone_2014,Carone_2015,Carone_2016,Zhang_2017,Haqq_Misra_2018,YangHZ_2019,Shields_2019,Eager_2020,Zhang_2020}, but they are not considered in this study. For example, air pressure can affect the absorption of infrared radiation via pressure broadening and collision-induced absorption, so air pressure can influence the radiative cooling rate and subsequently the overturning circulation.

In this study, we focus on the atmospheric circulation on planets holding Earth-like atmospheres (1 bar N$_2$ with trace gases) but without water vapor. However, the atmospheric compositions of rocky exoplanets are multifarious, depending on the activity of host star, surface outgassing, photochemistry, atmospheric collapse, atmospheric escape, runaway greenhouse, and other processes \citep{Toon_1980,Joshi_1997,Tian_2011,Ramirez_2014,Tian_2015,Owen_2016,Bolmont_2016,Bourrier_2017,Dong_2017,Dong_2018,Wordsworth_2018,Turbet_2018,Way_2019,Turbet_2020,Hu_2020}. Rocky planets may have a primordial H$_2$ atmosphere \citep{Pierrehumbert_2011,Genda_2017}, a thick H$_2$O-dominated atmosphere \citep{Turbet_2020}, an O$_2$-dominated atmosphere \citep{Luger_2015,Schaefer_2016}, or a CO$_2$-dominated atmosphere like Venus. Further work is required to consider whether our conclusions are applicable for various dry rocky planets with different atmospheric compositions.


\acknowledgments

We are grateful to the helpful discussions with Feng Ding, Ji Nie, Gang Chen, Xianyu Tan, and Daniel D.B. Koll. J.Y. acknowledges support from the National Natural Science Foundation of China (NSFC) under grants 42161144011 and 42075046.

\bibliography{sample63}{}
\bibliographystyle{aasjournal}



\end{document}